\documentclass[
               DIV=15, font size=10.5pt,
                 onecolumn,
               ]{scrartcl}  

\usepackage[dvipsnames]{xcolor}
\usepackage{psfrag} 
\usepackage{subfig}
\usepackage{hyperref}
\usepackage{amsmath}
\usepackage{graphics}
\usepackage{mathtools}
 \usepackage{pstool} 
\usepackage{todonotes}
\usepackage{gensymb}
\usepackage{bm}
\usepackage{multirow}
\usepackage{cancel}
\usepackage{float}
\usepackage{textcomp}
\usepackage{xcolor}
\usepackage[square, numbers,sort&compress]{natbib}
 
\usepackage{pgfplots}
\usepackage{pgfplotstable}
\usepgfplotslibrary{groupplots}
\usepackage{filecontents}
\usepackage{algorithm2e}
\usepackage{algorithmic}
\usepackage{siunitx}
\usepackage{tikz}
\usepackage{tikzscale}
\usepackage{tabularx} 

\usetikzlibrary{spy}
\usetikzlibrary{snakes,arrows,shapes}    
\usetikzlibrary{spy}
\usetikzlibrary{backgrounds}  
\usetikzlibrary{decorations}
\usetikzlibrary{positioning}
\usetikzlibrary{patterns}
\usetikzlibrary{calc} 
\usepackage{booktabs}   
\usepackage{makecell} 
\usepackage{colortbl}   

\usepackage{xspace}   
\usepackage{setspace}   
 
\usepackage{soul}
\usepackage{ulem}
\usepackage{currfile}
\usepackage{import}

\usepackage{authoraftertitle} % used for getting the authors 
\usepackage{authblk} % used for affiliations 
\usepackage{cleveref}

\pgfplotsset{compat=1.8}

\newcommand{\findmax}[3]{
    \pgfplotstablesort[sort key={#2},sort cmp={float >}]{\sorted}{#1}%
    \pgfplotstablegetelem{0}{#2}\of{\sorted}%
    \let #3=\pgfplotsretval%
}

\pgfplotscreateplotcyclelist{myCycleListBlackAndWhite}
{%
  {black, mark=o},
  {black, mark=square},
  {black, mark=triangle},
  {black, mark=diamond}, 
  {black, mark=pentagon}, 
  {black, mark=*},
  {black, mark=square*},
  {black, mark=triangle*},
  {black, mark=diamond*}, 
  {black, mark=pentagon*}, 
}

\definecolor{darkgreen}{rgb}{0,0.4,0} 
\definecolor{darkbrown}{rgb}{0.5, 0.396, 0.09}

\definecolor{c1}{rgb}{0.0, 0.4196078431372549, 0.6431372549019608}
\definecolor{c2}{rgb}{1.0, 0.5019607843137255, 0.054901960784313725}
\definecolor{c3}{rgb}{0.6705882352941176, 0.6705882352941176,
0.6705882352941176} \definecolor{c}{rgb}{0.34901960784313724, 0.34901960784313724, 0.34901960784313724}
\definecolor{c4}{rgb}{0.37254901960784315, 0.6196078431372549,
0.8196078431372549} \definecolor{c}{rgb}{0.7843137254901961, 0.3215686274509804, 0.0}
\definecolor{c5}{rgb}{0.5372549019607843, 0.5372549019607843,
0.5372549019607843} \definecolor{c}{rgb}{0.6352941176470588, 0.7843137254901961, 0.9254901960784314}
\definecolor{c6}{rgb}{1.0, 0.7372549019607844, 0.4745098039215686}
\definecolor{c7}{rgb}{0.8117647058823529, 0.8117647058823529,
0.8117647058823529}

\pgfplotscreateplotcyclelist{myCycleListColor}
{%
  {blue, mark=o},
  {red, mark=square},
  {darkbrown, mark=triangle},
  {darkgreen, mark=diamond},
  {black, mark=pentagon},
  {blue, mark=*},
  {red, mark=square*},
  {darkbrown, mark=triangle*},
  {darkgreen, mark=diamond*},
  {black, mark=pentagon*},
}

\pgfplotsset{every axis/.append style= 
              {
                font=\small,
                mark size=2,
                line width = 0.1,
                legend style={font=\small, mark size=3, draw=none, fill=none},
                legend cell align=left,
                cycle list name=myCycleListColor,
              }
            }
            
\pgfplotstableset
{
  every odd row/.style={before row={\rowcolor[gray]{0.9}}},
  every head row/.style=
  {
    before row=
    {%
      \toprule
    },
    after row=\midrule
  },
  every last row/.style=
  {
    after row=\bottomrule
  }
}

\newif\ifdrawboundingbox

\usetikzlibrary{external} % Is activated with the command
\tikzset{external/system call={pdflatex \tikzexternalcheckshellescape
-halt-on-error -interaction=batchmode -jobname "\image" "\texsource"}} 
\pgfkeys
{ 
  /pgf/number format/.cd,
  fixed,
  set thousands separator={\,},
  1000 sep in fractionals,
}

\newcolumntype{C}[1]{>{\centering\arraybackslash}m{#1}}
\newcolumntype{R}[1]{>{\raggedright\arraybackslash}m{#1}}
\newcolumntype{L}[1]{>{\raggedleft\arraybackslash}m{#1}}

\newcommand{\delete}[1]{\xspace}

\definecolor{Reviewer1}{rgb}{0.0, 0.0, 1.0}
\definecolor{Reviewer2}{rgb}{0.0, 0.5, 0.0}
\usepackage[dvipsnames]{xcolor}
\colorlet{Editor}{Black}
\colorlet{Reviewer1}{Black}
\colorlet{Reviewer2}{Black}
\colorlet{Reviewer3}{Black}
\colorlet{Reviewer4}{Black}

\title{\textcolor{Editor}{Image-based numerical characterization and experimental validation of tensile behavior of octet-truss lattice structures}}

\author[1]{N. Korshunova\thanks{\href{mailto:nina.korshunova@tum.de}{\texttt{nina.korshunova@tum.de}},
    Corresponding author}}
\author[2]{G. Alaimo}%\thanks{another.author@test.de}}
\author[3]{S. B. Hosseini}%\thanks{another.author@test.de}}
\author[2]{M. Carraturo}%\thanks{another.author@test.de}}
\author[2]{A. Reali}%\thanks{another.author@test.de}}
\author[3]{J. Niiranen}%\thanks{another.author@test.de}}
\author[2]{F. Auricchio}%\thanks{another.author@test.de}}
\author[4]{E. Rank}%\thanks{another.author@test.de}}
\author[1]{S. Kollmannsberger}%\thanks{another.author@test.de}}

 \affil[1]{Chair of Computational Modeling and Simulation,
 Technische Universit\"at M\"unchen, Germany}
 \affil[2]{Department of Civil Engineering and Architecture, University of Pavia, Italy}
 \affil[3]{Department of Civil Engineering, Aalto University, Finland}
 \affil[4]{Institute for Advanced Study, Technische Universit\"at M\"unchen, Germany}

\newcommand{\journal}{Additive Manufacturing}
\newcommand{\publicationDate}{March 07, 2021}

\usepackage{fancyhdr}
\date{}
\fancypagestyle{plain}{%
  \fancyhf{}%
  \fancyfoot[L]{
  \vspace{-1.25cm} 
  \footnotesize{Accepted to publication in \textit{\journal{}}  \hfill
  \publicationDate
  \\
  }} }

\crefname{figure}{Fig.}{Fig.}
\crefname{equation}{Eq.}{Eq.}
\crefname{table}{Tab.}{Tab.}

\drawboundingboxtrue
\drawboundingboxfalse 

\newcommand*{\figref}[2][]{%
	\hyperref[{fig:#2}]{%
		Fig.~\ref*{fig:#2}%
		\ifx\\#1\\%
		\else
		\,#1%
		\fi
	}%
}
\definecolor{changes}{RGB}{0,0,0}
\definecolor{changez}{RGB}{0,0,0}

\begin{document}  

\normalem

\maketitle  

%% Abstract ---------------------------------------
\vspace{-1.5cm} 
\hrule 
\section*{Abstract}
{
 The production of lightweight metal lattice structures has received much attention due to the recent developments in additive manufacturing (AM). The design flexibility comes, however, with the complexity of the underlying physics. In fact, metal additive manufacturing introduces process-induced geometrical defects that mainly result in deviations of the effective geometry from the nominal one. \textcolor{Editor}{This change in the final printed shape is the primary cause of the gap between the as-designed and as-manufactured mechanical behavior of AM products.} Thus, the possibility to incorporate the precise manufactured geometries into the computational analysis is crucial for the quality and performance assessment of the final parts. Computed tomography (CT) is an accurate method for the acquisition of the manufactured shape. However, it is often not feasible to integrate the CT-based geometrical information into the traditional computational analysis due to the complexity of the meshing procedure for such high-resolution geometrical models and the prohibitive numerical costs. In this work, an embedded numerical framework is applied to efficiently simulate and compare the mechanical behavior of as-designed to as-manufactured octet-truss lattice structures. \textcolor{Editor}{The parts are produced using laser powder bed fusion (LPBF).} Employing an immersed boundary method, namely the Finite Cell Method (FCM), we perform direct numerical simulations (DNS) and first-order numerical homogenization analysis of a tensile test for a 3D printed octet-truss structure. Numerical results based on CT scan (as-manufactured geometry) show an excellent agreement with experimental measurements, whereas both DNS and first-order numerical homogenization performed directly on the 3D virtual model (as-designed geometry) of the structure show a significant deviation from experimental data.
}
 \vspace{.2cm} 
%% Keywords ---------------------------------------
\vspace{0.25cm}\\
\noindent \textit{Keywords:} additive manufacturing, geometrical defects, octet-truss lattice, Finite Cell Method, computed tomography, numerical homogenization, Finite Element method
 
\vspace{0.35cm}
\hrule 
\vspace{0.15cm}
\captionsetup[figure]{labelfont={bf},name={Fig.},labelsep=colon}
\captionsetup[table]{labelfont={bf},name={Tab.},labelsep=colon}

\tableofcontents
\vspace{0.5cm}
\hrule 
%% Actual Content ---------------------------------
	\section{Introduction}
{
	Additive manufacturing (AM) has enabled the production of highly complex metal structures~\cite{Buchanan2019, Ngo2018, YangBook2017}. \textcolor{Editor}{In particular, the group of laser powder bed fusion processes (LPBF) allowed producing lightweight, strong metal parts with complex microstructure~\cite{ASTM52900}.} Lattice structures are a classic example of such sophisticated components that are difficult to manufacture by traditional methods~\cite{Maconachie2019}. Nevertheless, they are attractive for industrial applications due to their lightweight and their exceptional properties (e.g.~\cite{Shidid2016, Zhou2019}).
	
	However, AM design freedom comes with the difficulty of controlling the underlying physics. \textcolor{Reviewer2}{The process parameters highly influence the geometry at the sub-millimeter scale of the produced lattices. The field of these defects' characterizations have received significant attention in the past years (see, e.g.,~\cite{Maconachie2019,Liu2017,Melancon2017, Mazur2017, Bagheri2017}). In particular, the size of the melt pool can affect the lattice strut thickness, its shape, or connectivity~\cite{Dallago2019a, Bagheri2017}. Another imperfection that often occurs in the manufactured parts is the surface roughness caused by the attachment of the unmelted powder to the struts~\cite{Dong2017, Yan2012}. Often the internal porosity can occur in the final specimens. Furthermore, the material accumulation on the inclined and overhanging parts of the lattice are commonly observed, as the size of the melt pool strongly depends on the fraction of solid and powder parts in the layer~\cite{Mazur2017}. Overall, all geometrical imperfections} resulting from the \textcolor{Editor}{LPBF} process lead to a deviation in lattice components' expected mechanical performance. Although the optimization of the process parameters can guarantee a reduction of the gap between the as-manufactured and as-designed mechanical properties of the components, the process-induced defects are often not negligible~\cite{Maconachie2019}.  
	
	There is a considerable amount of research directed to the characterization of the mechanical behavior of additively manufactured lattice structures  (e.g.~\cite{Campoli2013,Dallago2019a,Dong2019,Liu2017,Pasini2019,Portela2018, Rashed2016, Tancogne2018, Vigliotti2014}). The goal of such investigations is to compare the behavior of as-manufactured and as-designed components and to characterize the influence of the process-induced defects on the behavior of the final parts. However, efficient incorporation of as-manufactured geometries into the computational simulation remains a challenge. \Cref{fig::workflow} provides a graphical overview of these approaches.

	To obtain an as-manufactured geometry for numerical modeling, both optical and scanning electron microscopy can be used~\cite{Platek2020, Saedi2018, Sanae2019, Sing2018}. However, these technologies only provide observable surface information. Yet, the lattice structures can exhibit large geometrical deviations in regions inaccessible by surface measurements. Such a deviation can be critical for the mechanical behavior of the final parts. These aspects make computed tomography (CT) a preferable choice for the acquisition of the as-manufactured geometry~\cite{Duplesis2018,Duplesis2020}. Recent advances in the Finite Element Analysis on CT-based geometries have shown that the numerical simulation of the as-manufactured parts leads to a better prediction of the mechanical behavior of the final AM components (e.g~\cite{Dallago2019a, Geng2019, Lei2019, Liu2017, Lozanovski2019, Wang2019}). There are two main approaches to incorporate the geometrical information from the CT scans into the numerical analysis: full 3D model reconstruction from the obtained CT data (e.g~\cite{Dallago2019a,Heinze2017, Geng2019, Wang2019}) and statistically equivalent CAD model generation (e.g~\cite{ Lei2019, Liu2017, Lozanovski2019}). Both methods encounter three main challenges. First, the reconstruction of a 3D model from a CT scan is a quite demanding task since it requires a lot of manual effort prior to the analysis \cite{Dallago2019a}. Second, a boundary conforming mesh, required for the numerical analysis, is not trivial to generate. Finally, due to the complex geometrical features of lattices, a numerical analysis of these structures is generally computationally expensive. Thus, for these kinds of structures, numerical homogenization methods are often preferred to a direct numerical simulation (DNS) of the considered experimental setup. To further reduce computational costs, beam elements are often used (e.g~\cite{Dallago2019a, Lei2019, Liu2017}) even though the 3D solid models clearly remain the most accurate.

	In the light of the above literature review, the main novelty of the present contribution is to provide an experimentally validated and effective computational workflow to numerically characterize the mechanical behavior of lattice structures directly from CT scan data images. The presented workflow requires neither a 3D model reconstruction of CT scan data nor a conforming mesh generation process, which are two main issues of most of the current numerical techniques employed to investigate mechanical properties of AM components.
	
	The present work aims at comparing the as-manufactured and as-designed octet-truss structures by using the DNS and first-order homogenization. To this end, the Finite Cell Method (FCM) is used directly on the CT images, omitting the step of a geometric model reconstruction. FCM provides an efficient tool to incorporate the process-induced geometrical defects of the manufactured lattice components into numerical analysis allowing the comparison of as-manufactured versus as-designed mechanical behavior. Additionally, this embedded technology provides an accurate prediction of the desired mechanical properties via a natural image-to-material-characterization workflow. 

	\textcolor{Reviewer2}{The tensile behavior of lattice structures usually represents a challenging task for experimental investigations. Compression tests are generally preferred due to their practical applicability to the lattices. Nevertheless, a large amount of research appears in this direction. In particular, the effective elastic properties of multiple lattice unit cells are discussed in~\cite{Refai2020a, Refai2020b}. Another significant amount of research is concerned with the investigation of the fatigue behavior of lattice structures~\cite{Dallago2017,Dallago2018a,Dallago2021,Lietaert2018}. However, most of the research focuses either on incorporating a specific set of geometrical defects into the numerical analysis into an ideal CAD model or limits the computational domain to a unit cell or an RVE. To the authors' knowledge, incorporating the as-manufactured geometries as a whole part considering a complex set of the arising imperfections is still challenging.} To fill this research gap, we start by providing a comprehensive experimental and numerical analysis of the tensile behavior of octet-truss lattice specimens. In particular, we numerically and experimentally investigate the tensile behavior of steel octet-truss lattices, comparing FCM results based on as-manufactured geometry to classical boundary conforming finite element analysis (FEA) conducted on the as-designed geometry, i.e., the original CAD model of the lattice structure.

	The present article is organized as follows. \Cref{sec:NumericsFCM} gives a summary of the high-order Finite Cell Method for image-based material characterization. In~\cref{sec:homogenization}, we further recall the fundamentals of the first-order homogenization for CT-based and CAD-based representative volume elements (RVE). \Cref{sec:manufacturing} provides the geometrical characteristics and description of the manufacturing process of the octet-truss lattice specimens. In~\cref{sec:experimentalTest}, we detail the experimental setup used for tensile testing of the produced lattice specimens. Then, the numerical findings are presented in~\cref{sec:numericalInvestigation}. For a comprehensive analysis of octet-truss lattices, a geometrical comparison of as-manufactured and as-designed structures is performed in~\cref{subsec:geometricalComparison}. In~\cref{subsec:tensileTest} the experimental tensile test is numerically simulated employing four different approaches: CT-based DNS, CT-based first-order homogenization, CAD-based DNS, and CAD-based first-order homogenization. Finally, in \cref{sec:conclusion} we draw the main conclusion before an outlook is presented. 
	
	\begin{figure}[H]
	\centering
	\def\svgwidth{\textwidth}
	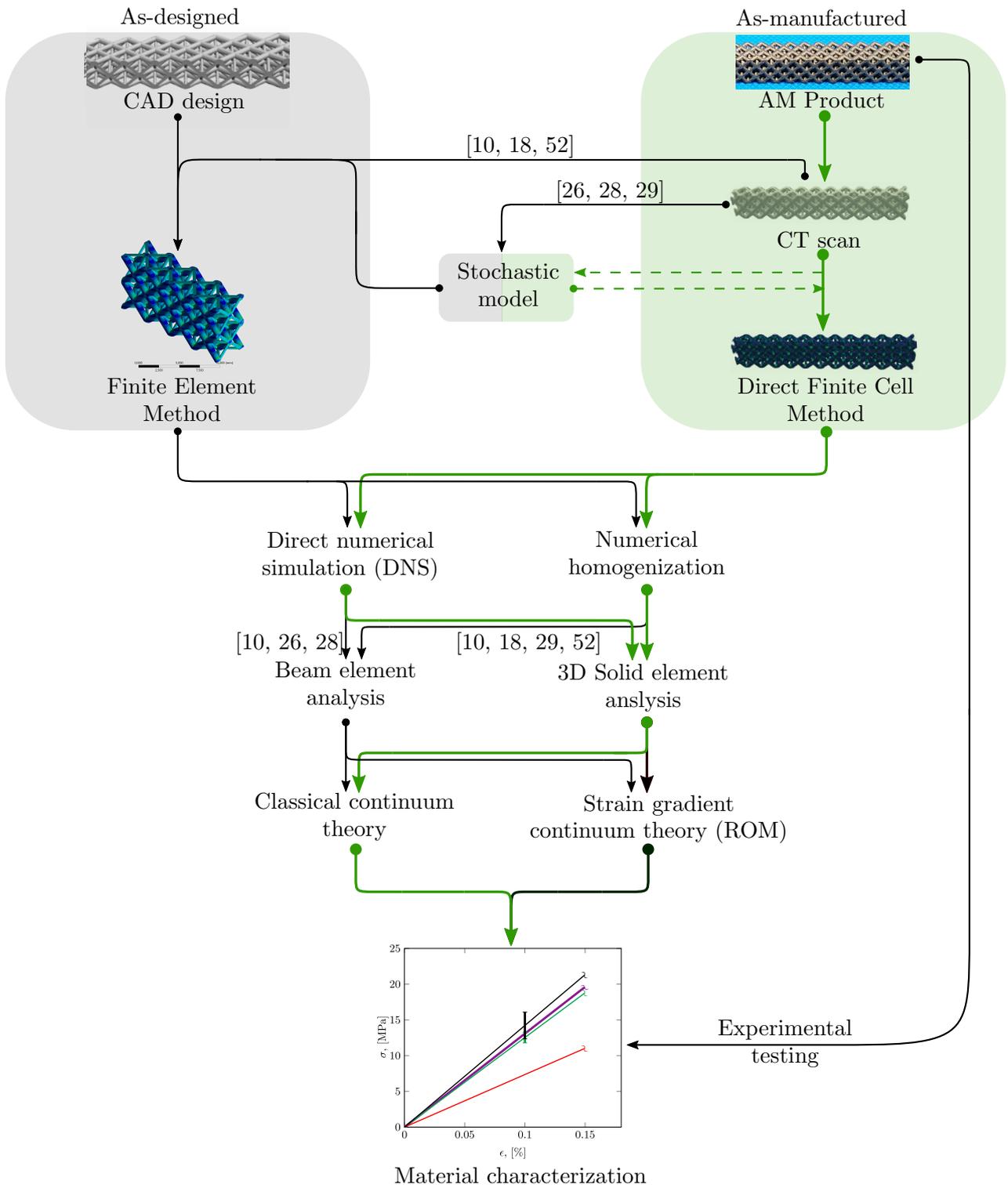
	\caption{Workflow for the incorporation of as-manufactured geometries into the computational simulation and comparison with as-designed simulation.}    
	\label{fig::workflow}
	\end{figure}
	
}
	\section{Material Characterization by the Finite Cell Method}
\label{sec:NumericsFCM}
{
	The geometrical models stemming from CT scans provide comprehensive information about as-manufactured components allowing for the characterization of the process-induced geometrical defects. The numerical simulation of CT-based microstructures is considered to be time-consuming and technically difficult due to the size of the considered data. FCM provides a powerful tool to efficiently simulate the geometrical models stemming directly from CT scan data images. The basic concept of this high-order immersed method is briefly summarized in the following, emphasizing the main advantages in the application to material characterization of AM products. For a more detailed description of FCM, the reader is referred to~\cite{Duster2017}.
	
	\begin{figure}[H]
		\centering
		\def\svgwidth{\textwidth}
		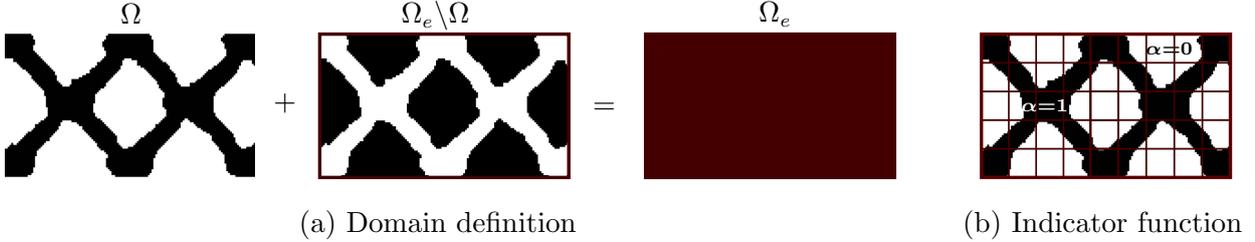
		\vspace*{0.2cm}
		\caption{Concept of the FCM (adapted from \citep{Duster2017}).}    
		\label{fig::FCMConcept}
	\end{figure}
	
	\textcolor{Reviewer1}{The main idea of the FCM is depicted in a simplified way on a two-dimensional slice of the octet-truss structure in~\cref{fig::FCMConcept}. The key concept} of FCM is to immerse the physical domain of interest $\Omega$ (see~\cref{fig::FCMConcept}) into a simply shaped fictitious domain such as a rectangle or a cube $\Omega_e\backslash \Omega$. Then, the union of these domains $\Omega_e$ can be discretized with a structured Cartesian mesh. The shape functions associated with this mesh can be chosen freely. In the present work, the integrated Legendre polynomials of degree $p$ are adopted \cite{Parvizian2007}. 
	
	Then, to recover the original mechanical problem, the weak form should be modified by an indicator function $\alpha(\bm{x})$,
	which can be formulated as follows:
	\begin{equation}
	\alpha(\bm{x})=\begin{cases}
	1 \qquad &\forall \bm{x}\in\Omega\\
	\approx 0 \qquad &\forall\bm{x}\in\Omega_e\backslash \Omega
	\end{cases}
	\label{eq:definitionAlpha}
	\end{equation}
	
	This definition of the indicator function ensures that the weak form defined on the immersed domain $\Omega_e$ is equivalent to the standard formulation in the energy sense up to a modeling error proportional to $\sqrt{\alpha}$. The linear elastic boundary value problem is then formulated as follows:
		\begin{equation}
	\text{Find}  \,\,\,  \bm{u} \in H^1_{\hat{u}}(\Omega_e)  \,\,\,  \text{such that}  \,\,\, 
	\mathcal{B}(\bm{v,u}) = \mathcal{F}(\bm{v}) \,\,\, \forall \bm{v} \in H^1_0(\Omega_e) 
	\label{eq::weakFormulationBVP}
	\end{equation}
	
		where the bi-linear and linear form can be written as:
		\begin{equation}
		\begin{aligned}
		\mathcal{B}(\bm{v,u}) &= \int\limits_{\Omega_e} \bm{\varepsilon(v)}:\alpha(\bm{x}) \bm{C(x)}:\bm{\varepsilon(u)} \, d\Omega_e + \beta_D\int_{\Gamma_D}\bm{v}\cdot\bm{u}\, d\Gamma_D \\
		\mathcal{F}(\bm{v}) &= \int\limits_{\Omega_e} \bm{v} \cdot \alpha(\bm{x})\bm{b}(\bm{x})\,d\Omega_e + \int\limits_{\Gamma_N} \bm{v} \cdot \hat{\bm{t}}\,d\Gamma_N + \beta_D\int\limits_{\Gamma_D} \bm{v} \cdot \hat{\bm{u}}\,d\Gamma_D
		\end{aligned}
		\label{eq:BilinearLinear}
		\end{equation}%
	In~\cref{eq::weakFormulationBVP} $H^1_{\hat{u}}$ indicates the boundary conforming first-order Sobolev space, $\Gamma_D$ is the part of the boundary of the physical domain $\Omega$, where Dirichlet boundary conditions $\hat{\bm{u}}$ are applied and $\Gamma_N$ corresponds to the Neumann boundary with the applied traction $\hat{\bm{t}}$. We recall that Dirichlet boundary conditions are enforced by the penalty method with a penalty parameter $\beta_D$~\cite{Babuska1973}.

	As the indicator function $\alpha(\bm{x})$ makes the integrands discontinuous, specially constructed integration algorithms must be applied for an accurate evaluation. Since the CT scans provide an underlying voxel structure with piecewise constant coefficients, a natural approach to construct a composed integration rule is to distribute integration points within every voxel in one finite cell (refer to~\cref{fig::IntegrationConcept}). Furthermore, as the structure of every element and the number of voxels per cell is regularly repeated throughout the whole structure, the domain integrals can be efficiently pre-computed before the numerical simulation~\cite{Yang2012a}. This approach drastically reduces the integration time allowing for efficient computation of complex problems based on CT scan data images.

	\begin{figure}[H]
		\centering
		\def\svgwidth{.7\textwidth}
		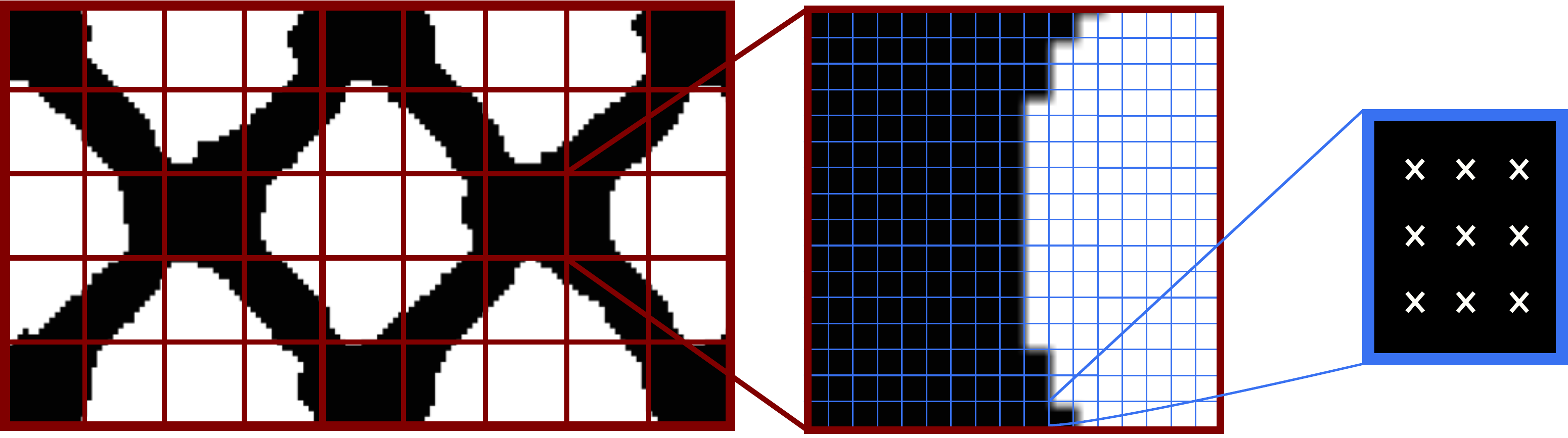
		\caption{Voxel-based pre-integration technique: left - the Finite Cell background mesh; middle - one Finite Cell with embedded voxels; right - integration point distribution within one voxel.}    
		\label{fig::IntegrationConcept}
	\end{figure}
	
Nevertheless, the common size of geometrical models stemming from CT scans is still very large due to the presence of small geometrical features. As an example, the images used for the further numerical investigations are of size $292\times176\times1768$ voxels, while the smallest features can be in the range of $3-7$ voxels.  Consequently, to solve the mechanical problem accurately, a high resolution of the numerical discretization is necessary. This, in turn, leads to a large size of numerical systems. To address this computational challenge, a hybrid parallelization of the high-order Finite Cell Method is used. The implementational details of this method are published in~\cite{Jomo2016, Jomo2019}. This approach allows performing numerical simulations on large CT scans capturing the mechanically significant small-scale defects.	

}

\section{First-order CT-based numerical homogenization}
\label{sec:homogenization}
{
	Although the proposed methods in~\cref{sec:NumericsFCM} enable numerical analysis on large CT scans with rather small geometrical features, the computational costs remain high. Thus, to further reduce the numerical size of the considered problems, we introduce the CT-based first-order numerical homogenization. This technique combined with the Finite Cell Method provides a computationally efficient alternative to evaluate the mechanical behavior in the linear regime. The two main advantages of this approach are the direct analysis of the representative volume elements (RVEs) stemming from CT images and the efficient pre-computation of the volume integrals in~\cref{eq:BilinearLinear} exploiting the underlying voxel structure. This allows for the flexible incorporation of as-manufactured geometries providing a reliable and efficient material characterization tool. In the following, only the main idea of numerical homogenization is recapitulated. For further details, the reader is referred to~\cite{Korshunova2020}. 
	
	The main concept of the CT-based numerical homogenization is visualized in~\cref{fig::HomogenizationConcept}. The complex macroscopic structure is split into RVE subdomains. These volumes must fulfill the length scale separation criteria; i.e., they should contain a sufficient number of heterogeneities to be representative of the overall macroscopic behavior. The RVEs are then treated as a material macroscopic point, which can be characterized by the effective material tensor $\bm{C}^*$.
	 
	\begin{figure}[H]
		\centering
		\def\svgwidth{.9\textwidth}
		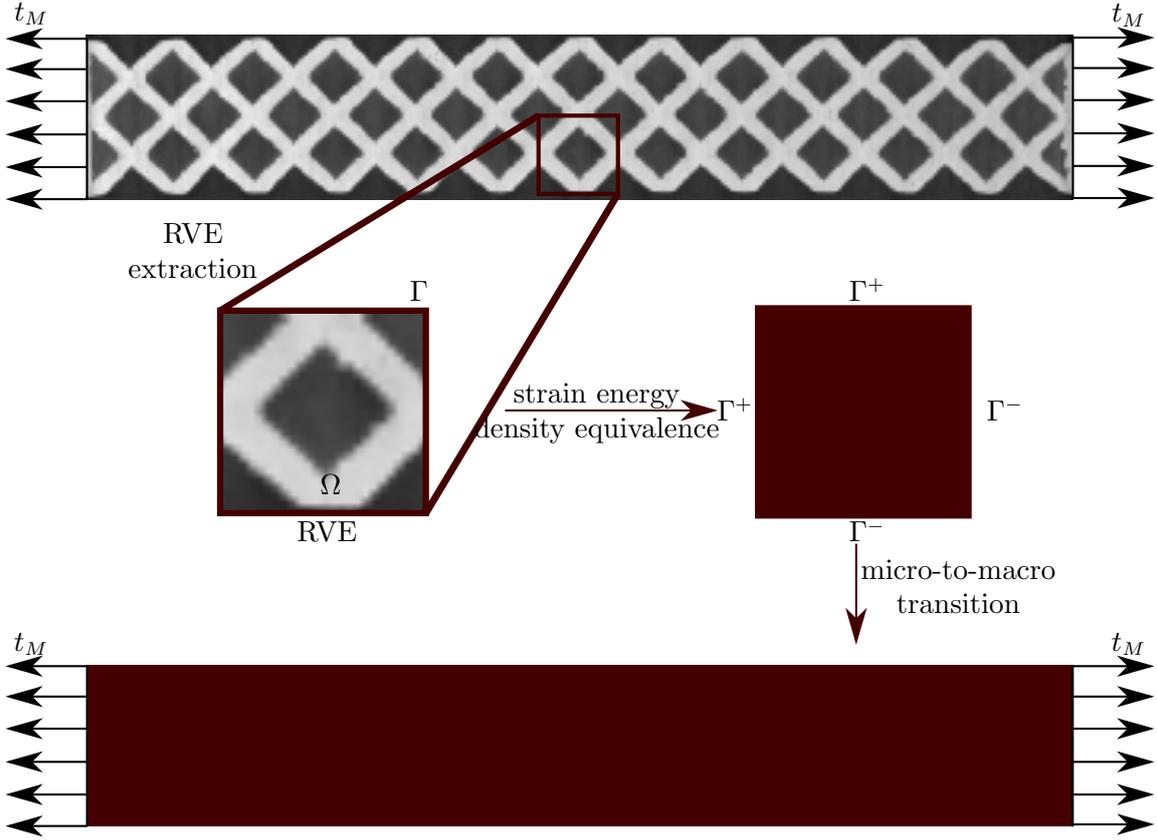
		\caption{The concept of the mean-field numerical homogenization.}    
		\label{fig::HomogenizationConcept}
	\end{figure}

	The first step in the homogenization procedure is RVE extraction. In this phase, the appropriate relation between the structure and the representative volume must be identified. As the RVE is considered to be a macroscopic material point, the macroscopic quantities are projected to the microscopic level by using the Hill-Mandel principle of the strain energy density equality~\cite{Gross2017,Nemat2013,Zohdi2004}:
	
	\begin{equation}
	\frac{1}{2}\left(\bm{\sigma}^M:\bm{\varepsilon}^M \right)= \int_{\Omega} \left( \bm{\sigma}:\bm{\varepsilon} \right)d\Omega
	\label{eq:HillMandelCondition}
	\end{equation}  
	where superscript $M$ indicates the macroscopic quantities and $\Omega$ is the RVE volume including all heterogeneities, e.g. voids or inclusions.
	
	After few mathematical manipulations the volume integral in~\cref{eq:HillMandelCondition} can be transformed to a boundary integral over $\Gamma$ which is the RVE boundary as indicated in~\cref{fig::HomogenizationConcept}:
	
	\begin{equation}
	\frac{1}{\Omega}\int_{d\Gamma}\left(\bm{u}-\left<\bm{\varepsilon}\right>_\Omega\bm{x}\right)\left(\bm{\sigma}-\left<\bm{\sigma}\right>_\Omega\right)\bm{n}d\Gamma = 0
	\label{eq:HillMandelConditionBoundary}
	\end{equation}  	
	where displacement, strain, and stress are the microscopic fields, the notation $\left<.\right>_{\Omega}$ indicates the averaging operator over domain $\Omega$, $\bm{x}$ is the coordinate vector, and $\bm{n}$ is the normal vector defined at the boundary $\Gamma$ of the RVE.
	
	The condition in~\cref{eq:HillMandelConditionBoundary} is a priori satisfied when the specific set of boundary conditions are applied to the RVE. These conditions commonly include the Kinematic Uniform Boundary Conditions (KUBC), Static Uniform Boundary Conditions (SUBC), and Periodic Boundary Conditions (PBC)~\cite{Pahr2008}. In the scope of this work, the latter is chosen for the numerical homogenization~\cite{Tian2019}. This is commonly preferred \textcolor{Reviewer4}{because} in the case of periodic structures, it delivers the exact effective material tensor. Also, for the microstructures which do not comply with the periodicity requirement, it is demonstrated to provide good estimates~\cite{Nguyen2012a}. \textcolor{Reviewer2}{Furthermore, the following order relation of the boundary conditions always hold when the size of the considered volume is smaller than RVE as follows~\cite{Hazanov1994, Suquet1985}:
	\begin{equation}
	C_{KUBC}\geq C_{PBC} \geq C_{SUBC}
	\end{equation}
	where $\geq$ indicates that the quadratic matrix forms of the difference between the material tensors	should be positive semi-definite.   
	}
	\textcolor{Reviewer2}{The examples considered in this paper are not strictly periodic. The unit cells, due to the manufacturing process, strongly differ from one another. Furthermore, very thin structures are considered. Thus, the Periodic Boundary conditions only deliver \textit{apparent} properties, i.e., represent an approximation of the effective behavior.}
	
	To impose the periodic boundary condition the boundary $\Gamma$ of the RVE is usually decomposed into two opposing parts  $\Gamma^+$ and  $\Gamma^-$ (see~\cref{fig::HomogenizationConcept}). Thus, each point with coordinates $\bm{x^+}$ on boundary $\Gamma^+$ is uniquely coupled with the point $\bm{x^-}$ on the opposite side $\Gamma^-$ via displacement and traction values as follows: 
	\begin{equation}
	\begin{aligned}
	\bm{u}(\bm{x^+})-\bm{u}(\bm{x^-}) &= \bm{\varepsilon}^M\Delta \bm{x}\\
	\bm{t}(\bm{x^+})&=-\bm{t}(\bm{x^-})
	\end{aligned}
	\label{eq:PBC}
	\end{equation}
	
	According to~\cite{Tian2019} the application of displacement constraints of~\cref{eq:PBC} can guarantee the uniqueness of the solution making unnecessary the application of the traction constraints in~\cref{eq:PBC}.

	Following this formulation, a periodic mesh must be constructed, ensuring paired nodes on mutually opposite sides of the RVE boundary $\Gamma^+$ and  $\Gamma^-$. Thus, when numerical homogenization is performed on non-periodic structures using the traditional Finite Element Method, the meshing procedure becomes rather cumbersome. In this contribution, the lattice structures under consideration are only approximately periodic. This means that every cell shows unique geometrical variations from a common underlying basic unit cell. Therefore, an application of periodic boundary conditions is challenging to impose in a traditional mesh-conforming discretization. However, the immersed Finite Cell Method as described in~\cref{sec:NumericsFCM} retains a structured Cartesian grid. This provides a unique possibility to efficiently apply the periodic boundary condition as the coupled nodes on the opposite sides of the RVE boundary are guaranteed by construction. This facilitates the application of the periodic boundary condition and diminishes the dependency on the underlying microstructure.

	The second step of the numerical homogenization is a micro-to-macro transition.  As the RVE represents a macroscopic material point, its fluctuating microscopic fields are required to be equal on average to the macroscopic quantities. Thus, the micro-to-macro relations can be formulated as follows:
	
	\begin{equation}
	\begin{aligned}
	\bm{\sigma}^{M}=\int_{\Omega}\bm{\sigma}d\Omega&=\frac{1}{\Omega}\int_{\Gamma}\bm{t}\otimes\bm{x}d\Gamma\\
	\bm{\varepsilon}^{M}=\int_{\Omega}\bm{\varepsilon}d\Omega&=\frac{1}{\Omega}\int_{\Gamma}\frac{1}{2}\left(\bm{u}\otimes\bm{n}+\bm{n}\otimes\bm{u}\right)d\Gamma
	\end{aligned}
	\label{eq:averaginRelations}
	\end{equation}	
	where $\otimes$ is the Kronecker product.

	Again, for periodic representative volumes with the voids crossing the boundaries, the integrals in~\cref{eq:averaginRelations} are straightforward to compute. Strictly speaking, the strain averaging in~\cref{eq:averaginRelations} is not even required as the Periodic Boundary Conditions ensure an a priori known macroscopic strain state. However, it is important to note that for the completeness of formulation, the strain fields in non-periodic volumes are ambiguous in the areas when the voids cross the boundaries. The Finite Cell Method, in turn, provides a solution to this ambiguity. As the indicator function $\alpha(\bm{x})$ defined in~\cref{eq:definitionAlpha} resembles the assumption of the void being a very soft material, it provides a unique definition of strains in the segments where the voids cross the boundary of the RVE and naturally preserves the consistency of strain energy.
	
	Finally, in the last step, the \textcolor{Reviewer2}{apparent} material tensor can be computed using the following relation:		
	\begin{equation}
	\bm{\sigma}^{M}=\bm{C}^*\bm{\varepsilon}^{M}
	\label{eq:effectiveMaterialTensor}
	\end{equation}
	Usually, no a priori assumptions about the type of macroscopic behavior are made. Thus, to make~\cref{eq:effectiveMaterialTensor} solvable, six linear independent load cases are constructed. The \textcolor{Reviewer2}{apparent} material tensor can, then, attain any material symmetry, e.g., isotropic, orthotropic, etc.. \textcolor{Reviewer2}{To further relax the periodicity requirement and account for the unit cells' variability in the structure, several volumes $n$ are extracted from the whole structure. }\textcolor{Reviewer3}{Then, every unit cell is homogenized which delivers an apparent material tensor $\bm{C}^*_{n}$. The whole structure's final response is then determined as a statistical average of all computed tensors together with the standard deviation, indicating the spread of the quantities in the whole domain. The statistical average can be computed in two ways. The first approach is to determine an element-wise mean value of the homogenized elasticity tensor. The second approach is to invert the computed effective material tensors, extract the directional effective Young's Modulus, and then average this quantity. The latter will be further used in this article.}

}

\newcommand{\pictureDir}{./sections/numerics/Pictures}
\section{Manufacturing of octet-truss lattices}
\label{sec:manufacturing}
{
	The numerical and experimental investigations are carried out on an octet-truss lattice structure. The unit cell is shown in~\cref{fig:unitCellOctet} with three orthogonal views indicating main dimensions. The overall cell size is $4\times4$~mm, while the horizontal strut thickness is $0.8$~mm, whereas the inclined struts are $0.4$~mm. This CAD model will be further used in numerical investigations and will be referred to as "as-designed" geometry.
	
	\begin{figure}[H]
	\centering
	\includegraphics[scale=0.25, trim={0cm 0.5cm 0cm 0.1cm}, clip]{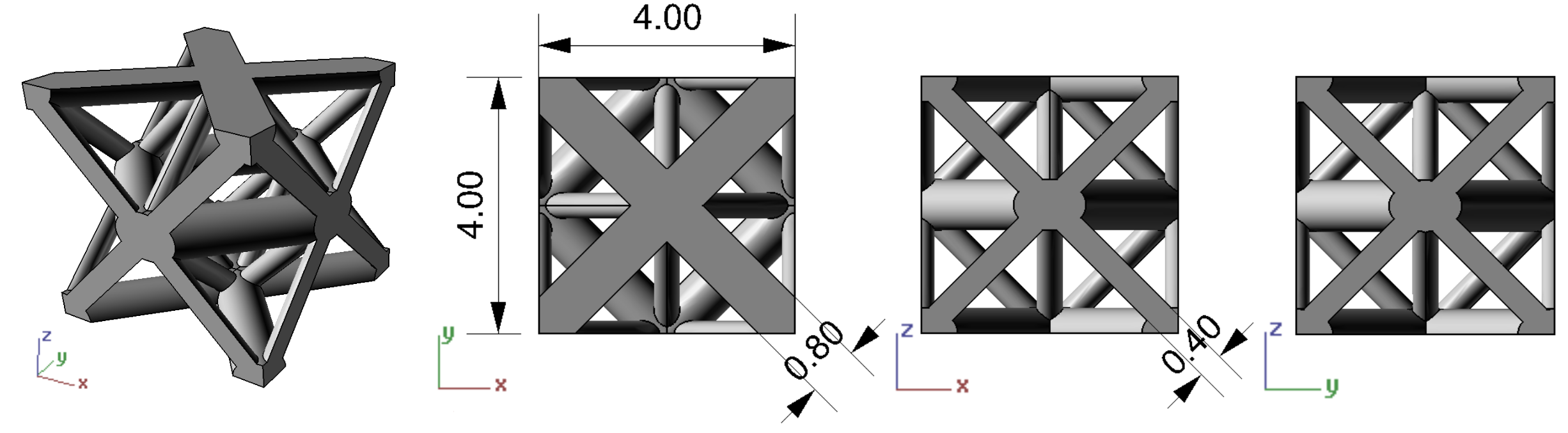}
	\caption{Octet-truss unit cell (dimensions are shown in mm).}
	\label{fig:unitCellOctet}
	\end{figure} 

	The specimens for experimental testing were produced \textcolor{Editor}{using the laser powder bed fusion} in the laboratory 3DMetal@UniPV\textcolor{Editor}{~\cite{ASTM52900}}. \textcolor{Editor}{The specimens were produced with} SS316L metal powder, using a Renishaw AM400 \textcolor{Editor}{LPBF} system.  The adopted \textcolor{Editor}{LPBF} process parameters are detailed in \cref{tab::ProcessParameters}; in particular, we use a 200 W laser power and a layer thickness of 50 $\mu$m.

		\begin{table}[H]
		\centering
		\caption{Process parameters for octet-truss manufacturing.}
		\label{tab::ProcessParameters}
		\begin{tabular}{|c|c|}
			\hline
			Process parameters &  Value \\\hline
			Build plate temperature & $170^{\circ}$C $\pm\, 1^{\circ}$C  \\
			Chamber temperature & $35^{\circ}$C $\pm\, 5^{\circ}$C  \\
			Layer thickness & $50\,\mu$m $\pm\, 1\,\mu$m  \\
			Hatch spacing & $110\, \mu$m $\pm\, 2\, \mu$m  \\
			Scan speed & $1200$ mm/s $\pm\, 2$ mm/s  \\
			Laser power & $200$ W $\pm\, 0.1$ W \\
			Laser spot size & $70\, \mu$m $\pm\, 1\, \mu$m \\\hline
		\end{tabular}
	\end{table}
 
	 According to the material data sheet provided by the powder producer~\cite{ManualReinshaw}, the considered setup leads to a bulk material with Young's modulus $190\, \text{GPa} \pm 10 \text{GPa}$ along the longitudinal direction with the density of $\rho=7.99 \, g/cm^3$. The produced specimens after heat treatment at $400^{\circ}C$ in the chamber Nabertherm LH120/12 for 2 hours are shown in~\cref{fig:PrintedSpecimens}.

%\begin{figure}[H]
%	\centering
%	\includegraphics[scale=0.3]{\pictureDir/process.jpg}
%	\caption{Manufactured experimental specimens.}
%	\label{fig:manufacturingProcess}
%\end{figure} 

\newcommand{\graphDir}{./sections/experiments/Pictures}

\begin{figure}[H]
	\centering
	\includegraphics[scale=0.1,trim={45cm 8cm 17cm 18cm},clip]{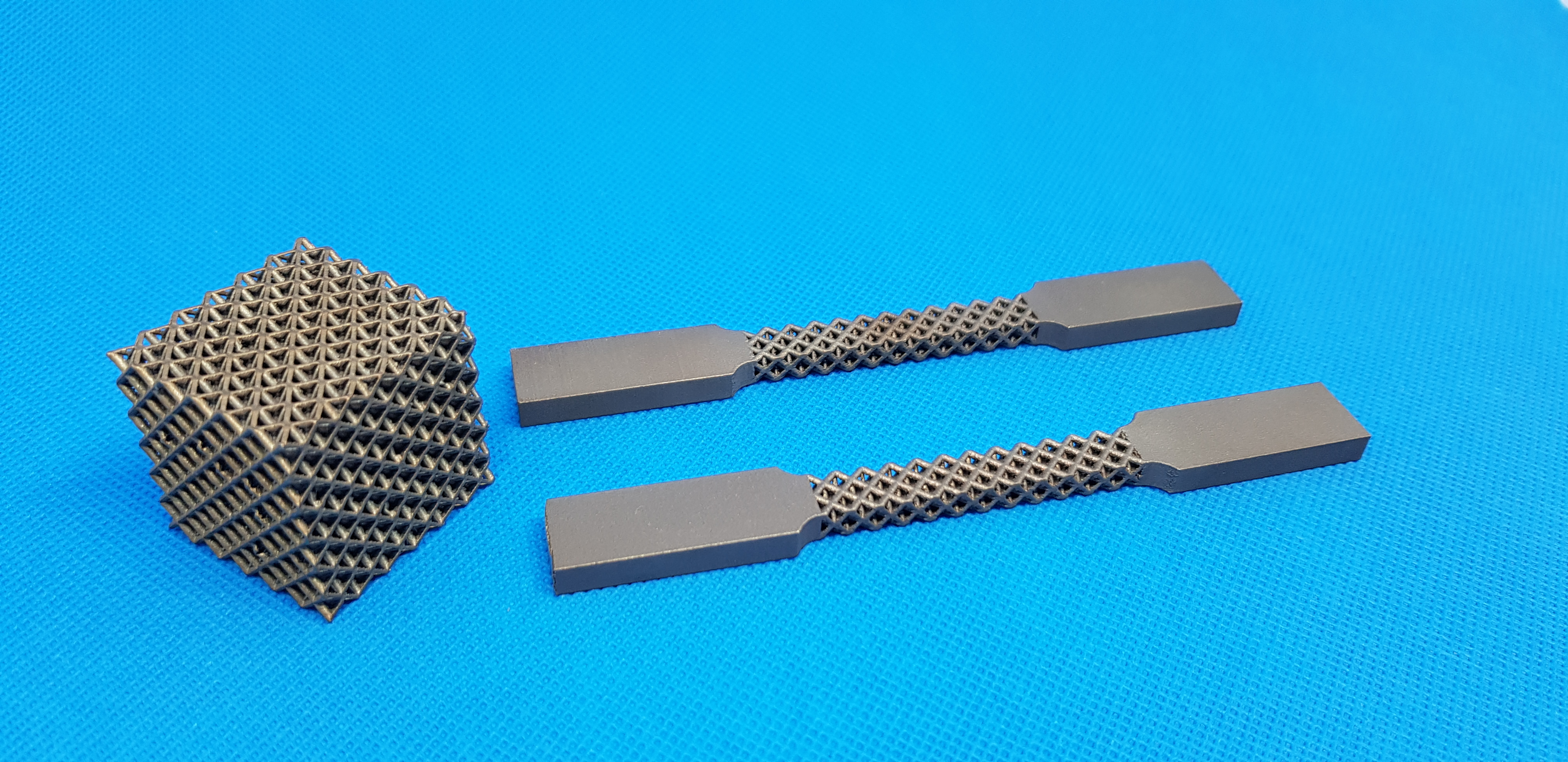}
	\caption{Printed tensile specimens after heat treatment.}
	\label{fig:PrintedSpecimens}
\end{figure}

	After printing, a tensile specimen was subjected to a computed tomography before experimental testing. The geometry acquisition was performed with the CT scanner Phoenix V resulting in a voxel resolution of 27 $\mu m$. These images deliver information about the internal structure of the printed specimens and will be further referred to as "as-manufactured" geometry. 

}
	\section{Experimental setup}
\label{sec:experimentalTest}
{

\newcommand{\graphDir}{./sections/experiments/Pictures}
	For a comprehensive experimental study of the printed specimens, two main investigations are undertaken. First, to characterize as-manufactured macroscopic geometrical variations, the overall porosity is measured. Next, the specimens are tested under tensile loading to evaluate the effective directional Young's modulus. 
	
	Commonly, the porosity is defined as a void fraction in the specimen over its total volume:
	\begin{equation}
	\phi = 1 - \frac{V_{m}}{V}
	\label{eq:porosity}
	\end{equation}
	For experimental porosity evaluation, four rectangular specimens with the same microstructure were produced with the same process parameters as in~\cref{tab::ProcessParameters}. The overall volume $V$ was then determined by measuring the dimensions of these parts. The volume occupied by the material $V_m$ was computed in two steps. First, the printed specimens were weighted to determine their mass. Then, the volume was calculated according to the bulk density specified by the powder producer~\cite{ManualReinshaw}. Therefore, the porosity was averaged over four considered specimens. The measurement uncertainty of the porosity values $\phi$ was estimated based on the accuracy of the used instrumentation.
	
	To experimentally validate the proposed numerical framework for the mechanical evaluation of lattice components, the tensile test was performed. For the aim of this work, only the elastic regime was investigated. Herein we provide a brief description of the experimental setup.

	The lattice specimen was tested under displacement control, at room temperature, on an MTS Insight test system, with computer control and data acquisition. The strain is measured with a video extensometer (ME-46 video extensometer, with 1 mm resolution and a camera field of view of 200 mm) at the mid-section of the specimen (gage section). Following ASTM E111 recommendations, the displacement rate is set to 2 mm/min, and it is selected to produce failure \textcolor{Reviewer3}{between 1 and 10 minutes}. The Young's Modulus was then computed according to ASTM E111 standard~\cite{ASTME111} with the corresponding measurement error.

}
	\section{Results and discussion}
\label{sec:numericalInvestigation}

	In this section, the results of the geometrical and numerical comparison on the as-manufactured and as-designed structures are discussed (see definition in~\cref{sec:manufacturing}). In~\cref{subsec:geometricalComparison}, these specimens are compared qualitatively via a visual identification of the overall geometrical variations and quantitatively by estimating the geometric characteristics, such as macroscopic porosity. Then, in~\cref{subsec:tensileTest} the mechanical behavior of the tensile specimen is analyzed in detail. Numerical simulations of the tensile test are performed on both as-manufactured and as-designed geometries and compared to experimental values. The former is analyzed employing FCM introduced in~\cref{sec:NumericsFCM}, whereas the latter using conforming mesh finite elements using the CAD geometry. \textcolor{Reviewer4}{All numerical investigations utilizing the FCM are performed using the in-house immersed high-order FEM code \textit{AdhoC++}. The code has been validated and verified on multiple benchmarks and industrial examples (see, e.g.,~\cite{Duster2017, Elhaddad2018, Elhaddad2015, Yang2012a}). } Furthermore, the first-order homogenization technique presented in \cref{sec:homogenization} is also employed to simulate the tensile test again using both as-manufactured and as-designed geometries and compare them to DNS results.

	\subsection{Geometrical comparison of as-designed and as-manufactured specimens}
	\label{subsec:geometricalComparison}
{	
	\newcommand{\graphDir}{./sections/experiments/Pictures}
	
	At first, the as-manufactured tensile specimen is compared to the as-designed CAD model used for its printing. The CT scan was converted to a point cloud and overlayed with the designed CAD model. \textcolor{Reviewer3}{To achieve the best possible fit of these two models, a two-step procedure is employed. First, a coarse alignment of the point cloud to a CAD model is performed. In this step, the model bounding boxes are aligned, followed by the point pairs picking. Then, the fine registration is run to achieve the best possible fit of the two models. The geometrical manipulations are performed with the open-source software CloudCompare.} The distance between the point cloud and the CAD model is computed as the nearest neighbor distance (see~\cref{fig:PrintedSpecimensComparison}). The specimen was printed in the $x-$direction shown in~\cref{fig:PrintedSpecimensComparison}. Furthermore, we can observe partially melted powder particles hanging on the diagonal struts (having axes out of the yz plane) as depicted in~\cref{fig:PrintedSpecimensComparison}c,d and, on the horizontal struts in~\cref{fig:PrintedSpecimensComparison2}. \textcolor{Reviewer2}{Quantitive characterization of a set of occurring geometrical imperfections as in~\cite{Dallago2017, Liu2017, Pasini2019} would greatly benefit this analysis. However, as the focus of this work is the mechanical comparison of as-designed and as-manufactured lattices, we do not consider a detailed statistical quantification in this paper.}	
	
	\begin{figure}[H]
		\captionsetup[subfigure]{labelformat=empty}
		\centering
		\subfloat[(a) Front view ]{\includegraphics[scale=0.31,trim={0cm 0cm 15cm 0cm},clip]{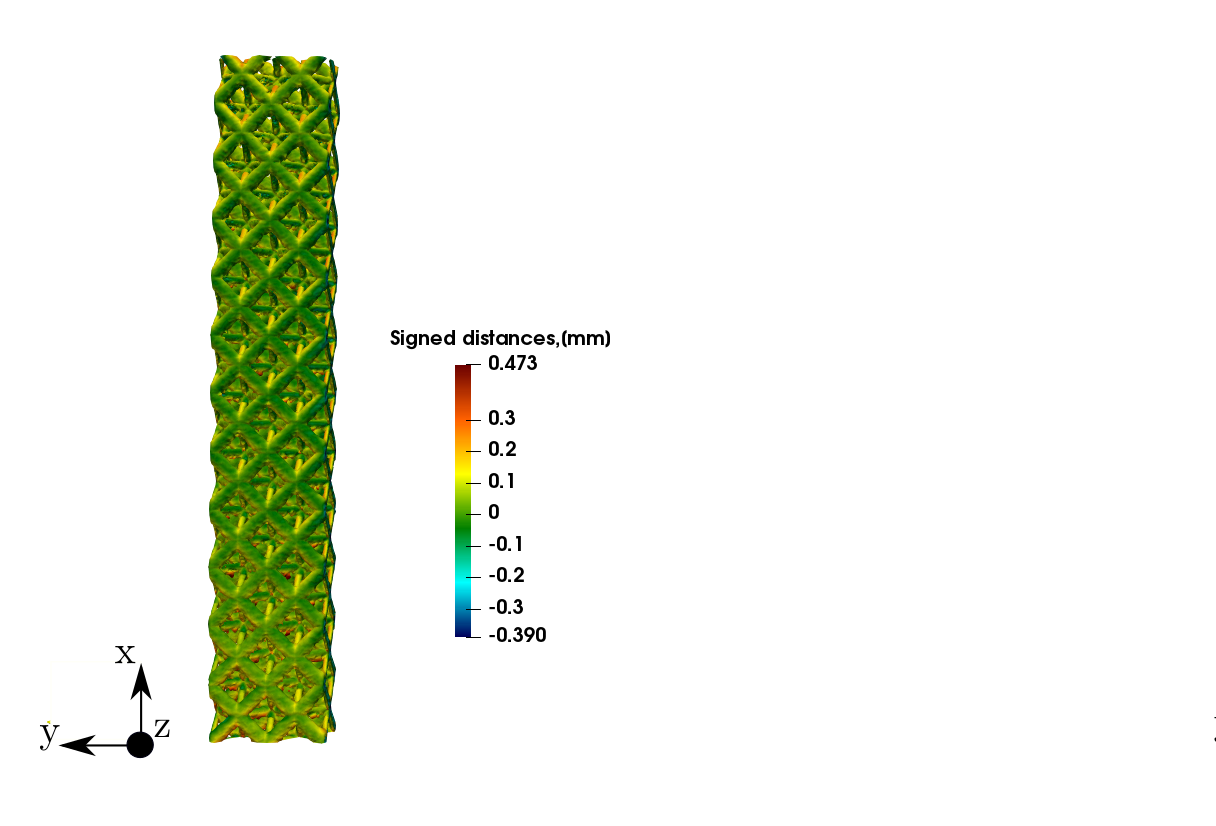}}
		\hspace*{1cm}
		\subfloat[(b) Back view ]{\includegraphics[scale=0.31,trim={0cm 0cm 15cm 0cm},clip]{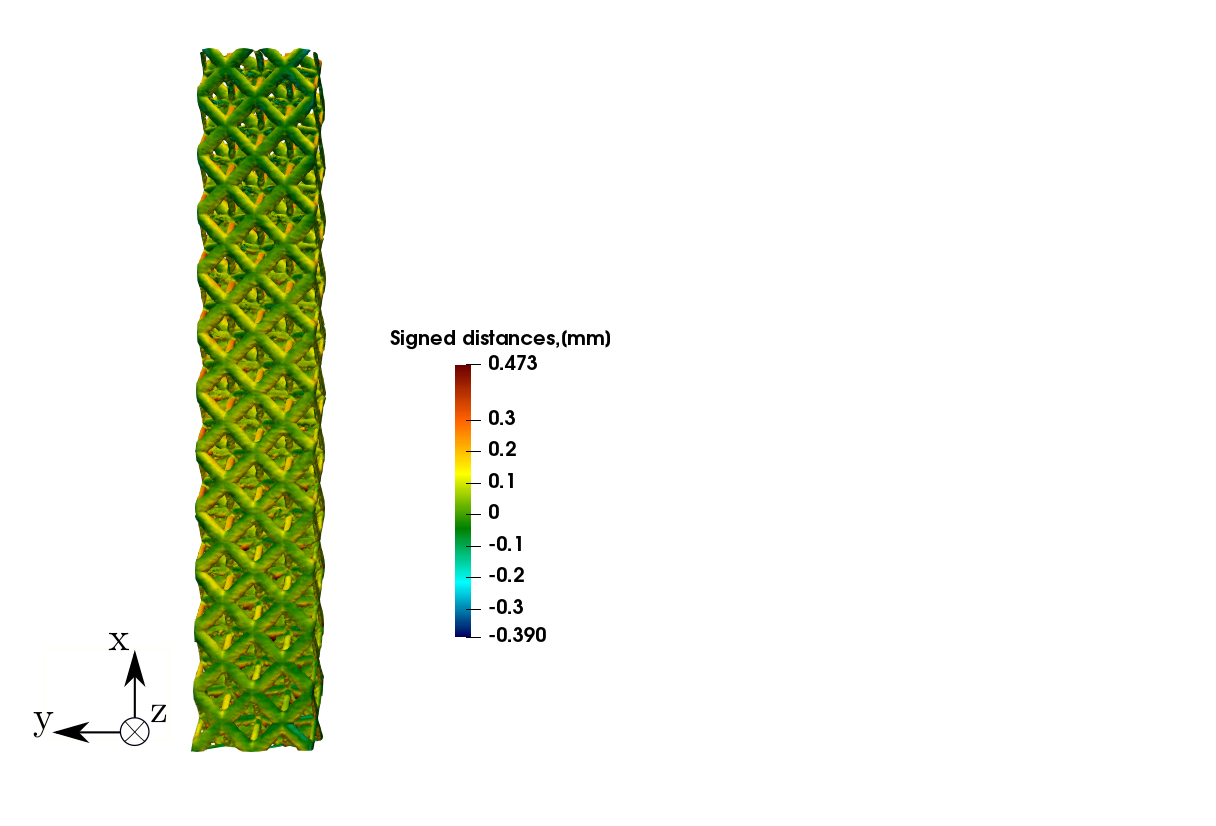}}\\
		\subfloat[(c) View from the bottom	]{\includegraphics[scale=0.23,trim={0cm 0cm 2cm 0cm},clip]{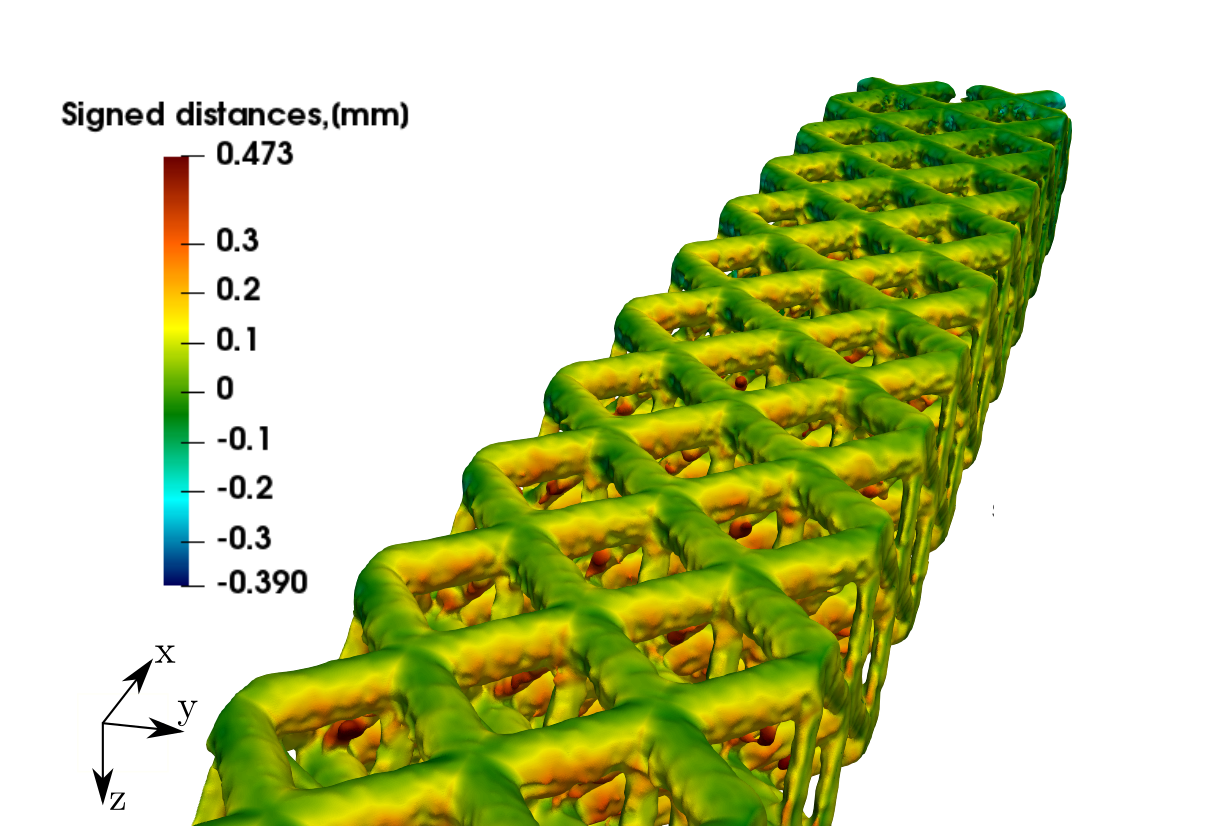}}
		\hspace*{1cm}
		\subfloat[(d) View from the top]{\includegraphics[scale=0.23,trim={0cm 0cm 2cm 0cm},clip]{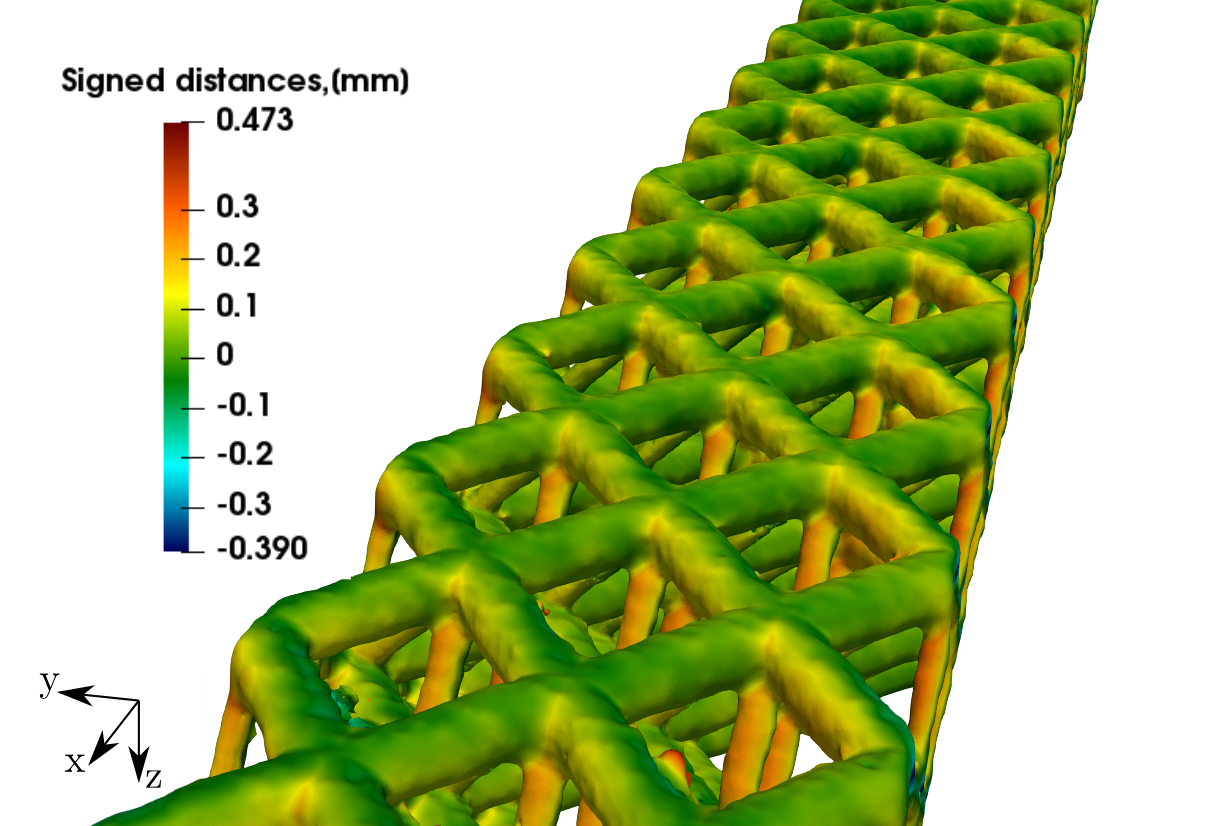}}\\
		\caption{Comparison of as-manufactured and as-designed tensile specimen.}
		\label{fig:PrintedSpecimensComparison}
	\end{figure}

	\begin{figure}[H]
		\centering
		\includegraphics[scale=0.2]{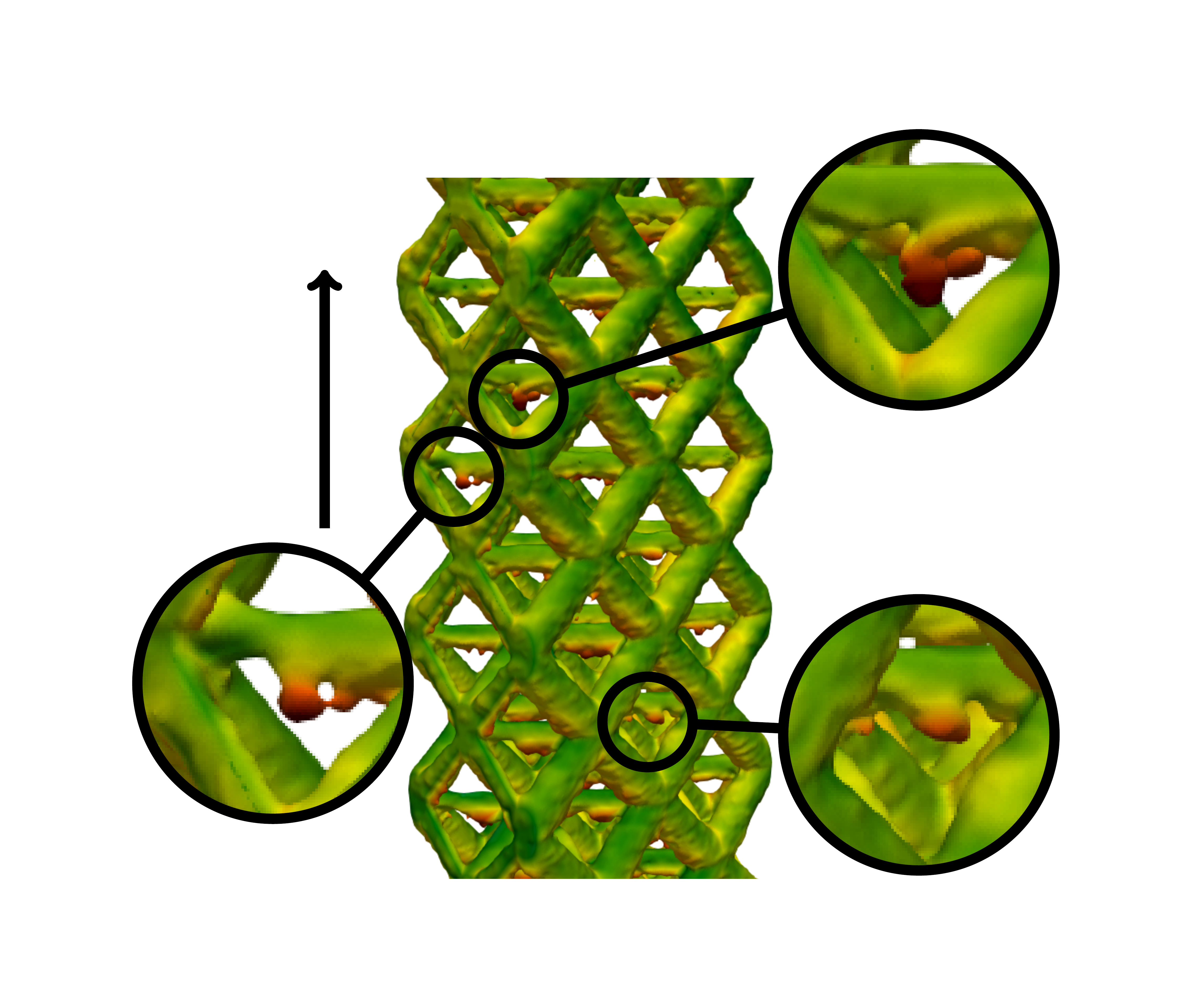}
		\caption{Partially melted powder grains attached to the horizontal struts in the octet-truss tensile specimen. The arrow is pointing in the build direction.}
		\label{fig:PrintedSpecimensComparison2}
	\end{figure}

	Then, the CT-based porosity is compared with the CAD-based and experimental values in~\cref{tab:PorosityTensileTogetherWithCADAndCT}. The experimental results, together with the estimated instrumentation error, are determined according to the procedure described in~\cref{sec:experimentalTest}. The CAD-based porosity is measured by using an as-designed geometrical model of the specimens as shown in~\cref{sec:manufacturing}. The CT-based results are obtained similarly using computed tomographic images of the corresponding as-manufactured specimens. 
	
		\begin{table}[H]
			\centering
			\caption{Comparison of the porosity of the tensile specimen.}
			\label{tab:PorosityTensileTogetherWithCADAndCT}
			\begin{tabular}{|c|c|c|c|}\hline
				Specimen &  CAD-based porosity, [-] & Experimental porosity, [-] & CT-based porosity, [-] \\\hline
				Octet-truss & 0.756 & 0.646 $\pm$ 0.002 &  0.668\\\hline
			\end{tabular}
		\end{table}

	Note that the CT-based porosity is in good agreement with the experimentally determined value. As-manufactured results deviate from the average experimental porosity by 3.4\%. As expected, CAD-based porosity is larger than the printed one. This finding is strongly supported by the geometrical comparison of the CAD and CT-based model of \cref{fig:PrintedSpecimensComparison2}. 
	
	One of the reasons why the as-designed parts have higher porosity values is that the struts with axes in the yz-plane are always larger in the printed components than in the designed ones. This is because partially melted powder particles remain attached to the surface of the manufactured component opposite to the build direction. Such defects are a well-known natural side-effect of the melting process together with the excess material collection in the nodes, where the higher temperatures occur. On the contrary, the CT-based porosity is in good agreement with the experimental values making us confident that the chosen CT scan resolution is accurate enough to obtain a reliable geometry representation.
	 
}
	
\newcommand{\graphDir}{./sections/results/Pictures}

\subsection{Tensile test of an octet-truss lattice structure}
\label{subsec:tensileTest}		
	
\paragraph{{ CT-based direct numerical simulation of a tensile test \newline}}

\begin{figure}[H]
		\centering
		\includegraphics[scale=0.15,trim={0cm 12cm 0cm 20cm},clip]{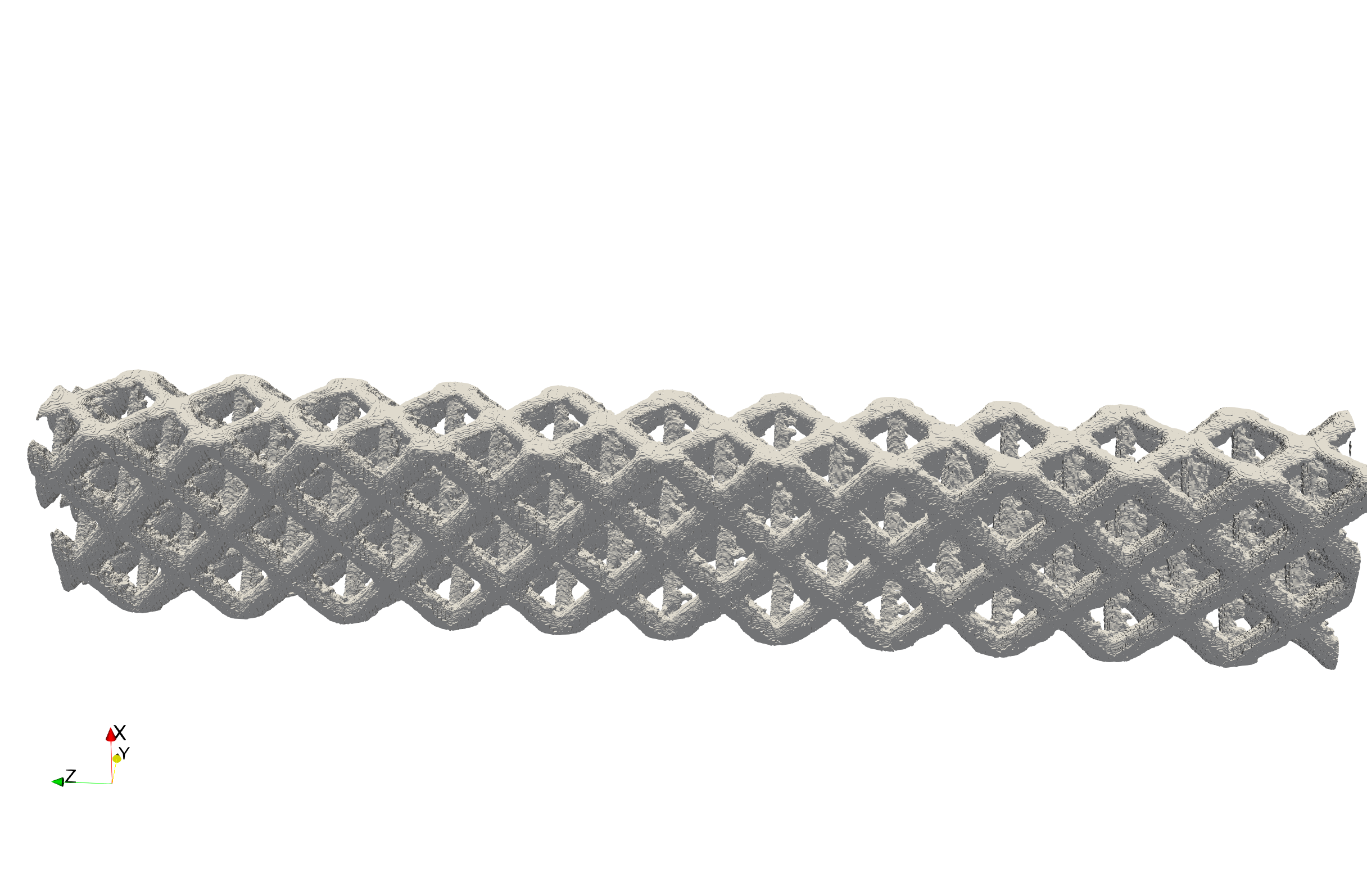}
		\caption{CT scan data image of the middle part of the tensile specimen.}
		\label{fig::CTScan}
	\end{figure}

	Herein, the computed tomography of the tensile specimen is investigated numerically by using FCM as described in~\cref{sec:NumericsFCM}. The middle part of the obtained CT scan is presented in~\cref{fig::CTScan}. To correctly compute the numerical results, we need to accurately evaluate the value of the spatial scalar function $\alpha(\bm{x})$ defined in~\cref{eq:definitionAlpha}. In the present work - as the material of the product is metal - the indicator function can be directly deduced from the CT scan since the contrast between material and void is very high. To distinguish among the holes where \textcolor{Reviewer4}{ the indicator function $\alpha(\bm{x})$ is set to $10^{-11}$ for further numerical investigations} and the metal where $\alpha(\bm{x})= 1$, the threshold level of the Hounsfield units is set to be $14\,500$.  \textcolor{Reviewer3}{The standard single thresholding technique was sufficient in this case and resulted in the model with the porosity indicated in~\cref{tab:PorosityTensileTogetherWithCADAndCT}.} All gray values above the limit are considered to be material, while everything below is classified as void.

\begin{figure}[H]
		\captionsetup[subfigure]{labelformat=empty}
		\centering
		\subfloat[(a) Displacement field along the build direction ($z$-axis) in mm]{\includegraphics[scale=0.15,trim={0cm 0cm 0cm 20cm},clip]{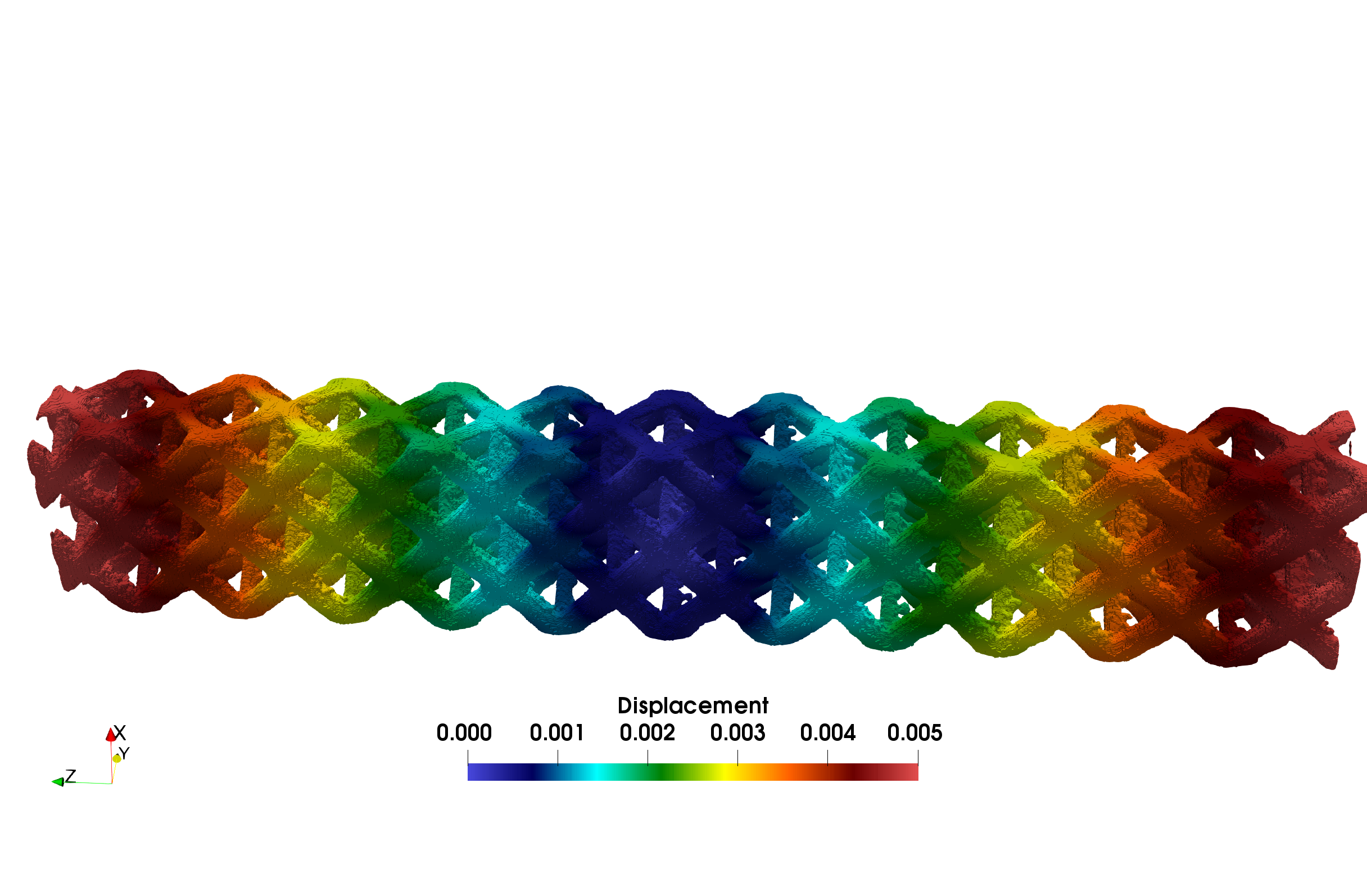}}
		\\
		\subfloat[(b) Von Mises stress distribution in MPa]{\includegraphics[scale=0.15,trim={0cm 0cm 0cm 20cm},clip]{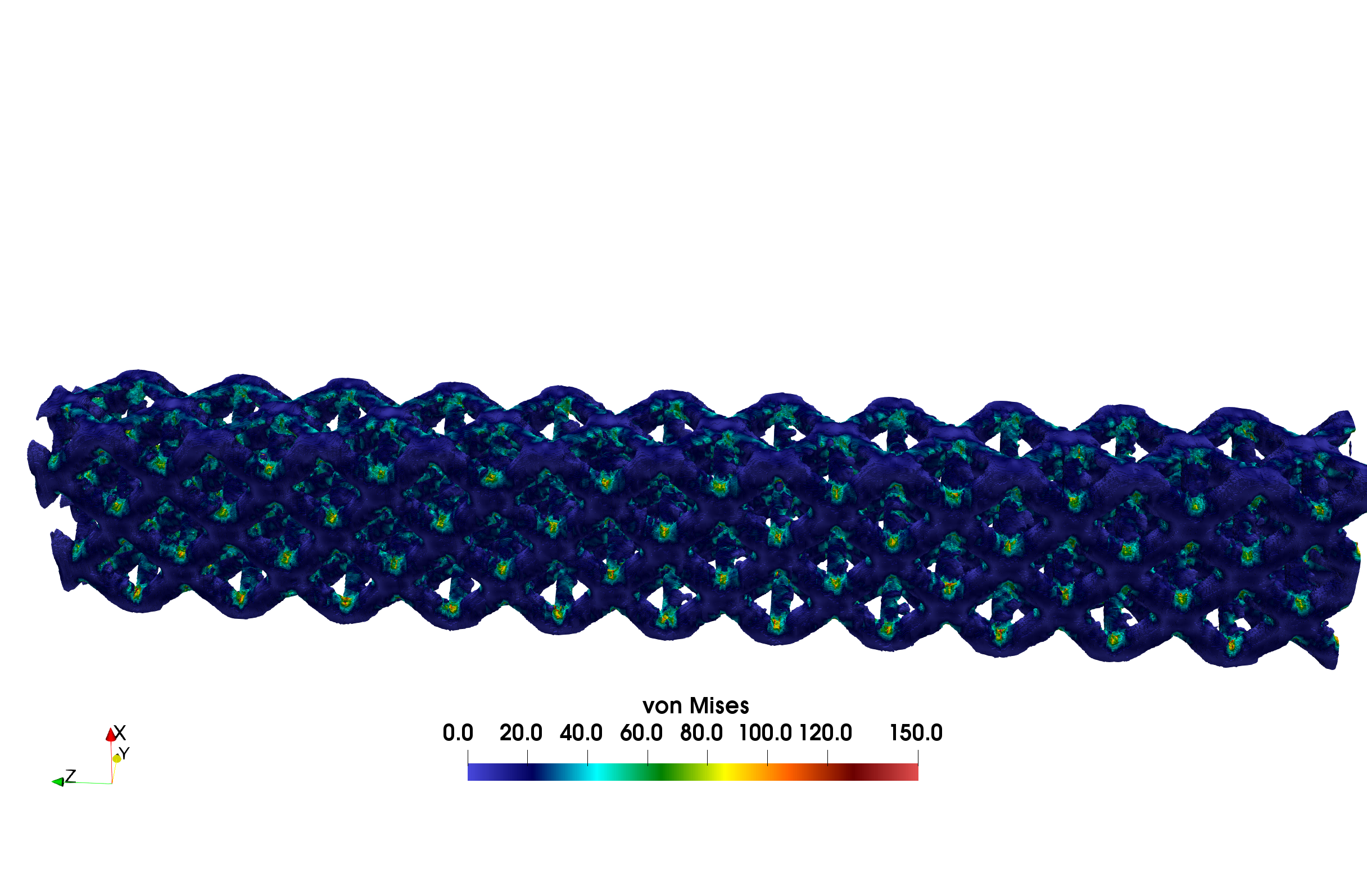}}
		\caption{CT-based numerical analysis: Representative displacement and stress fields in the middle part of the tensile specimen.}
		\label{fig::CTBasedDisplacementStress}
	\end{figure}

	Having defined the indicator function, a convergence study of the effective Young's modulus is performed. \textcolor{Reviewer3}{As mentioned in~\cref{sec:experimentalTest}, this work focuses on the linear elastic characteristics of the octet-truss lattices. Thus, the linear elastic material model is used for all numerical investigations.} The results of this analysis are summarized in~\cref{fig:ConvergenceDNS}. Given two different FCM discretizations with $73\times 44\times 442$ and $146\times 88\times 884$ (i.e., embedding $4\times 4\times 4$ and $2\times 2\times 2$ voxels in one finite cell, respectively), a convergence study is carried out, increasing the polynomial order of the approximation space. An example of the displacement and stress field arising in such specimen is shown in~\cref{fig::CTBasedDisplacementStress}. In the following, we refer to these numerical results obtained directly computing a tensile experiment on the full specimen as CT-based DNS results. The relative error is determined with respect to an overkill numerical resolution, where every voxel is an element with the polynomial degree of $p=2$ resulting in $390\,112\,737$ DOFs. The reference solution, in this case, has a value of $12\,736$ MPa. \textcolor{Reviewer4}{The necessary computational resources indicated in~\cref{tab:ComputationalResources} are very large, i.e., this overkill result was computed on the SuperMUC cluster of TUM. Due to large size of the linear system, this computation is memory expensive.}
	
	The final results for the CT-based numerical simulations are shown together with the experimental and CAD-based values in~\cref{tab::YoungsModulusNumerics}. As a final estimate, Young's modulus of $13\,081$ MPa is chosen. This value corresponds to a Finite Cell discretization of $146 \times 88 \times 884 $ cells of polynomial degree $p=3$ resulting in $98\,316\,435$ DOFs. \textcolor{Reviewer4}{The necessary computational resources from the CoolMUC cluster of TUM are summarized in~\cref{tab:ComputationalResources}. Compared to the Voxel-FEM solution, the size of the linear systems is much smaller, thus it does not require a large memory storage.} The final numerical value is 4.4\% different from the experimental value and within the limits of the estimated instrumentation error. Moreover, the results of these analyses show a clear convergence trend of the numerical solution, and the method presents a robust behavior even for coarse and low-order discretizations. The above observation confirms the possibility of adopting such an approach for material characterization of as-manufactured AM lattice structures.
	\begin{table}[H]
		\centering
		\begin{tabular}{|c|c|c|c|}
			\hline			
			\textcolor{Reviewer4}{Setup} & 
			\textcolor{Reviewer4}{Computational nodes} & 
			\textcolor{Reviewer4}{CPU per node} & 
			\textcolor{Reviewer4}{Wall-clock time} \\
			\hline
			
			\textcolor{Reviewer4}{DNS FCM} & 
			\textcolor{Reviewer4}{40} & 
			\textcolor{Reviewer4}{28} & 
			\textcolor{Reviewer4}{52 min}\\
			
			\textcolor{Reviewer4}{DNS Voxel-FEM$^\ddagger$}& 
			\textcolor{Reviewer4}{90} &
			\textcolor{Reviewer4}{28} & 
			\textcolor{Reviewer4}{22 min}\\
			\hline
		\end{tabular}
	\caption{
		\textcolor{Reviewer4}{Comparison of necessary computational resources for a direct numerical simulation on full specimen ($^\ddagger$ - memory extensive computation).}}
	\label{tab:ComputationalResources}
	\end{table}

	\newcommand{\tikzDir}{./sections/results/Pictures/TikZ/FullTensileTest/graphs}
\newcommand{\dataDir}{./sections/results/Pictures/TikZ/FullTensileTest/data}
\begin{figure}[H]
	\centering
	\hspace*{-1.5cm}
	\begin{minipage}{.5\textwidth}
		\centering
		\includegraphics[height=0.4\textheight]{\tikzDir/convergenceFull.tikz}
		\caption{Convergence of the effective Young's modulus $E^*$.}    
		\label{fig:ConvergenceDNS}
	\end{minipage}    
	\renewcommand{\tikzDir}{./sections/results/Pictures/TikZ/HomCTScan/graphs}
	\renewcommand{\dataDir}{./sections/results/Pictures/TikZ/HomCTScan/data} 
	\begin{minipage}{.5\textwidth}
		\centering
		\includegraphics[height=0.4\textheight]{\tikzDir/stressStrain.tikz}
		\caption{Stress-strain curve of the octet-truss lattice structure.}    
		\label{fig:allTestStressStrain}
	\end{minipage}    
\end{figure}   

\paragraph{{CT-based numerical homogenization results \newline}}
	Due to the high computational costs of the DNS numerical analysis, we decide to investigate less demanding (at least from a computational point of view) approaches. The first-order homogenization technique described in~\cref{sec:homogenization} is employed. The results of all tests are shown together in the stress-strain graph in~\cref{fig:allTestStressStrain}. Due to the high randomness of the as-manufactured geometry, homogenization is performed on \textcolor{Reviewer3}{$n=24$ unit cells as described in~\cref{sec:homogenization}} and the mean value and standard deviation are indicated with the black line in~\cref{fig:allTestStressStrain}. The deviation from the mean value is computed by considering the spread of the homogenized Young's modulus through the computed unit cells. \textcolor{Reviewer4}{An example of the displacement and stress distribution in one of the as-manufactured unit cells is shown in~\cref{fig::CTBasedDisplacementStressHomogenization}.}

	\textcolor{Reviewer4}{
	\begin{figure}[H]
		\captionsetup[subfigure]{labelformat=empty}
		\centering
		\subfloat[	\textcolor{Reviewer4}{(a) Displacement field for the load case in $x-$direction}]{\includegraphics[scale=0.15,trim={0cm 0cm 0cm 0cm},clip]{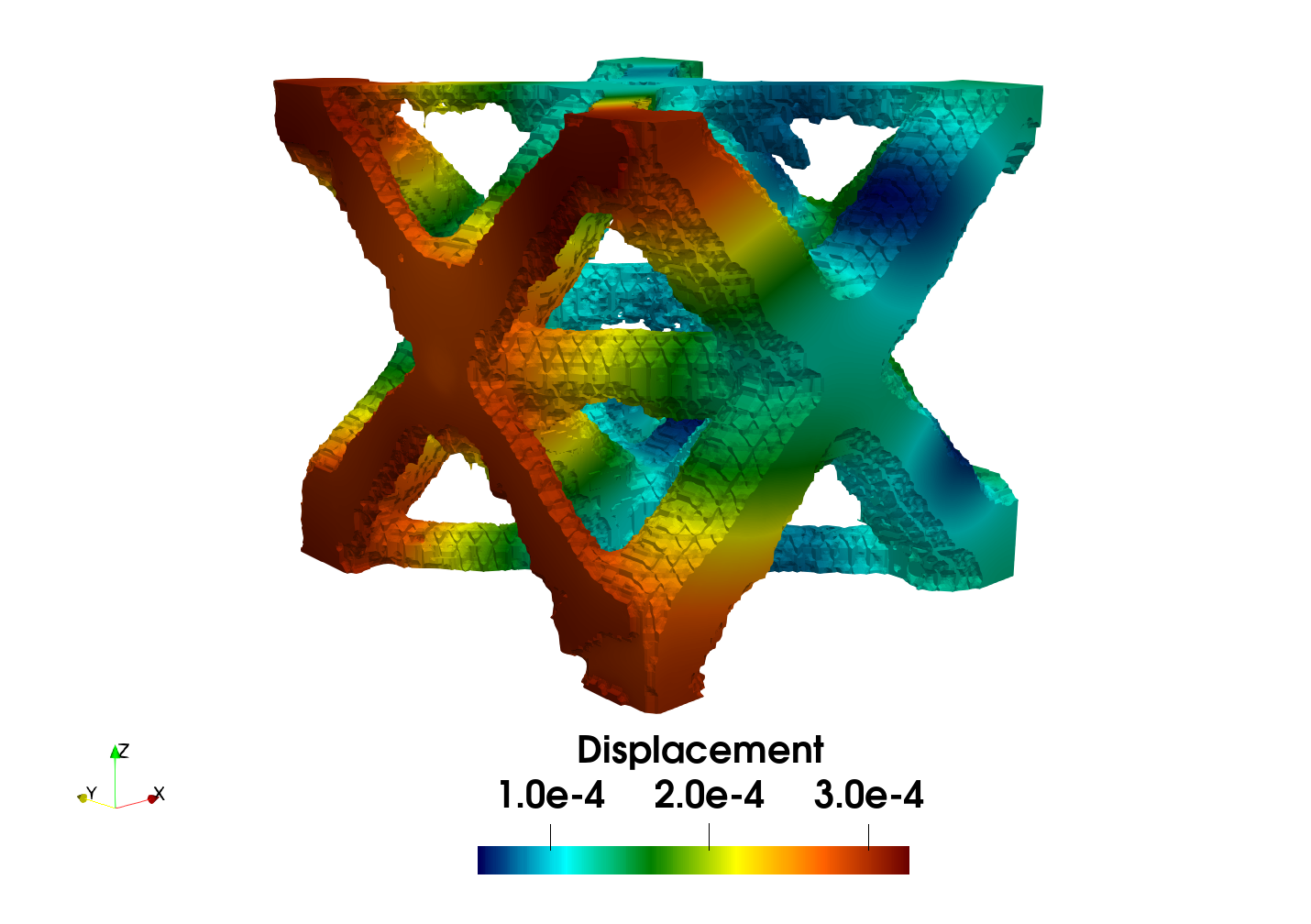}}
		\subfloat[	\textcolor{Reviewer4}{(b) Von Mises stress distribution in MPa}]{\includegraphics[scale=0.15,trim={0cm 0cm 0cm 0cm},clip]{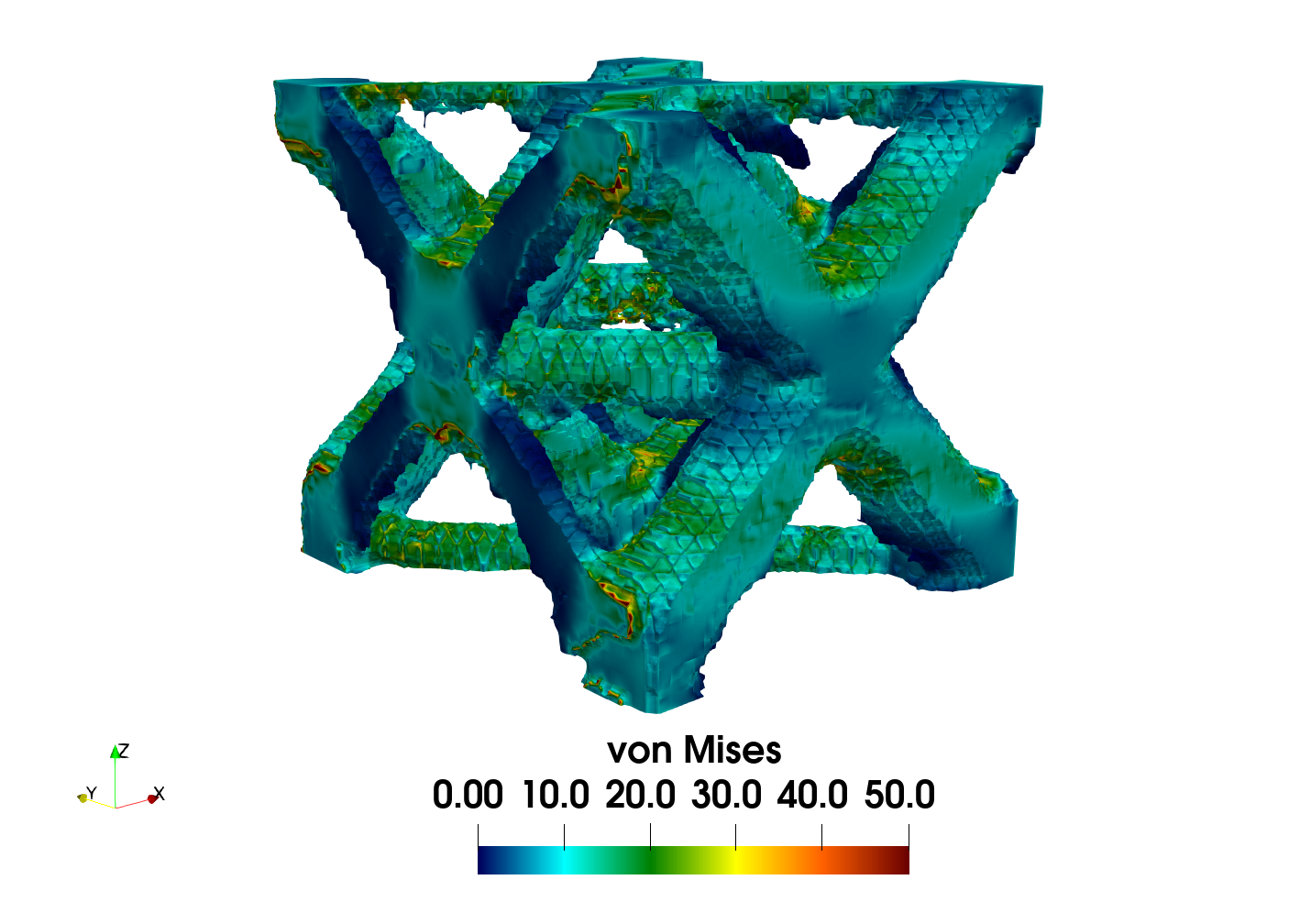}}\\
		\subfloat[	\textcolor{Reviewer4}{(c) Displacement field for the load case in $yz-$plane}]{\includegraphics[scale=0.15,trim={0cm 0cm 0cm 0cm},clip]{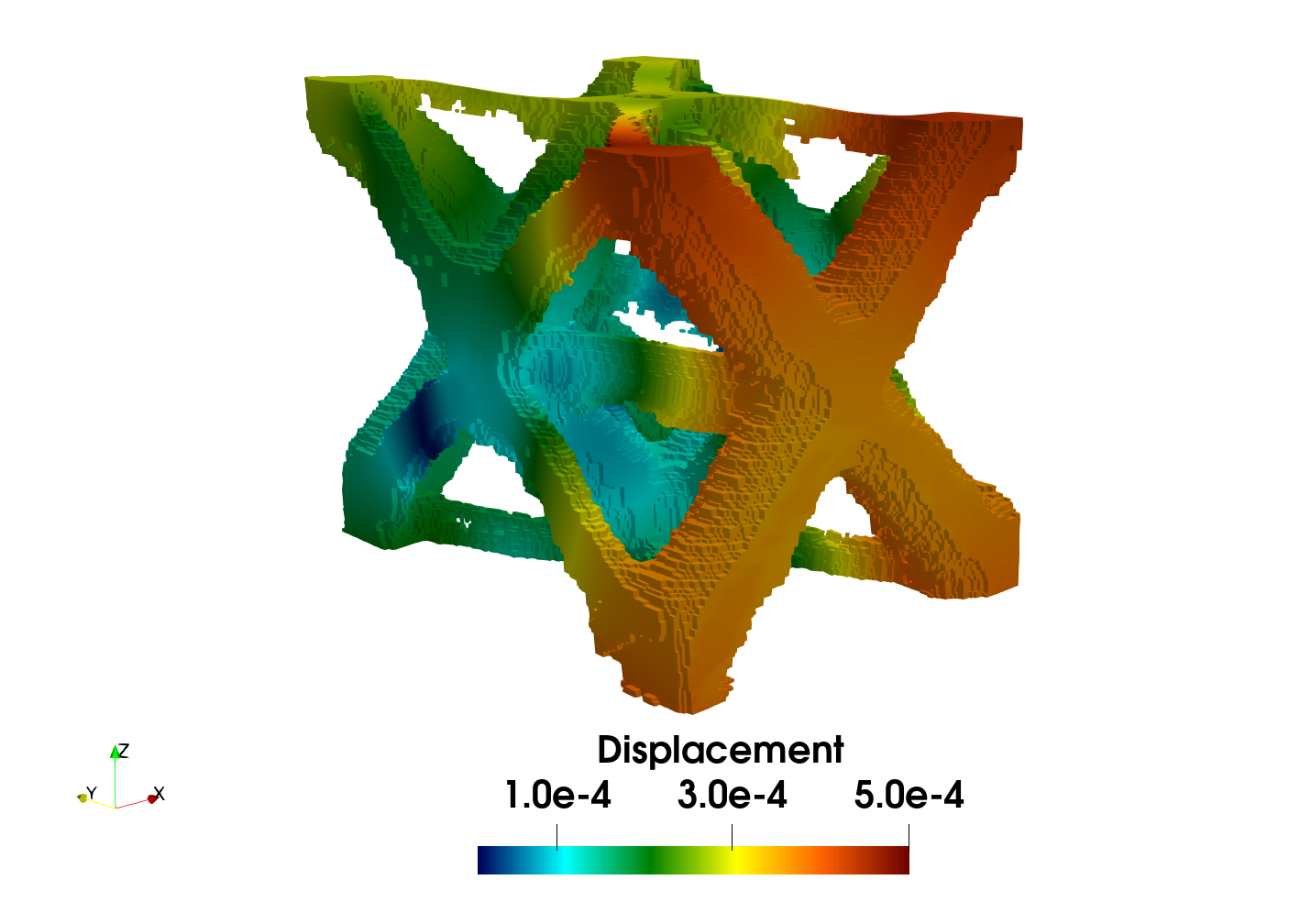}}
		\subfloat[	\textcolor{Reviewer4}{(d) Von Mises stress distribution in MPa}]{\includegraphics[scale=0.15,trim={0cm 0cm 0cm 0cm},clip]{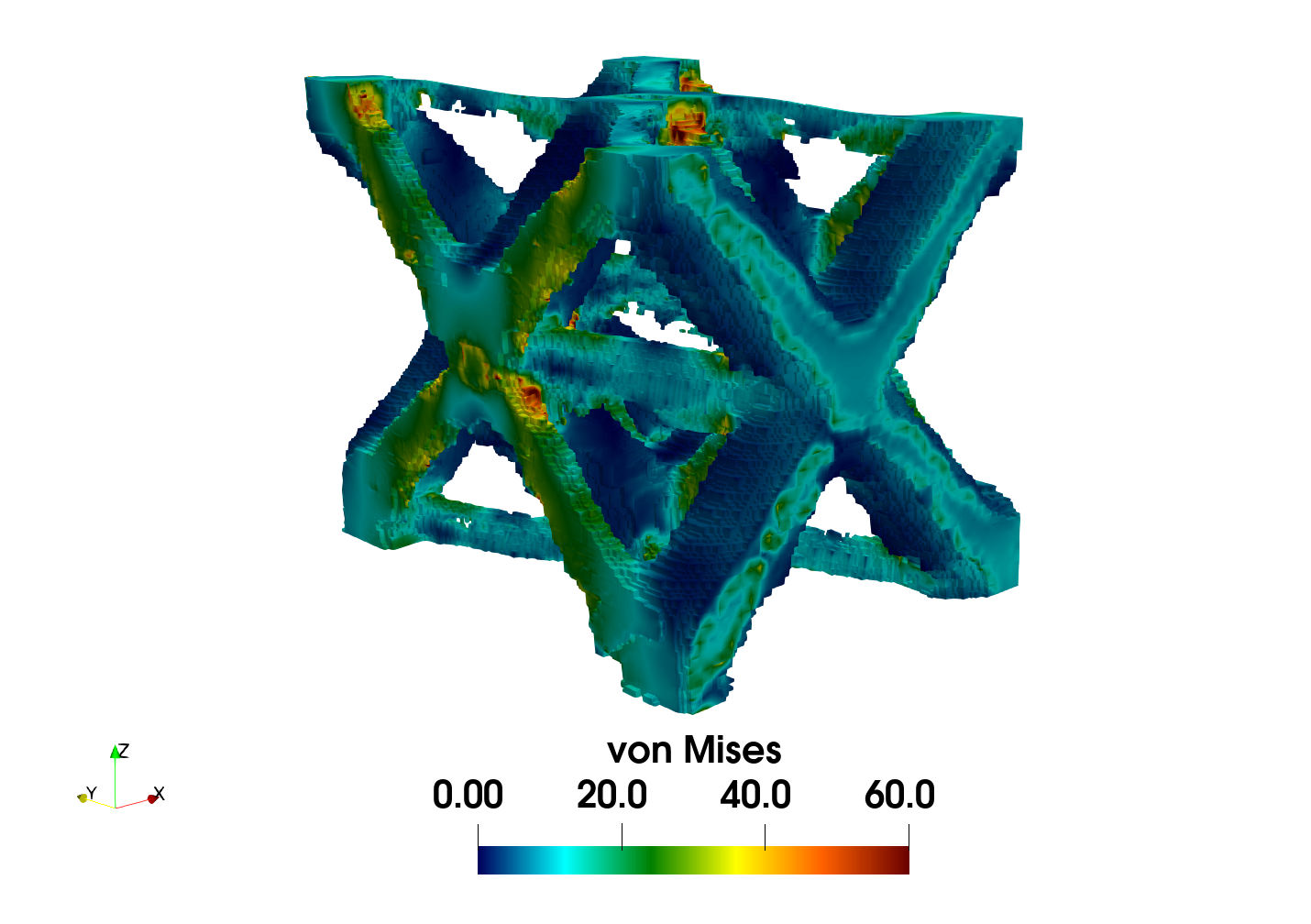}}
		\caption{	\textcolor{Reviewer4}{CT-based numerical homogenization: Representative displacement and stress fields on one unit cell for tensile and shear load case.}}
		\label{fig::CTBasedDisplacementStressHomogenization}
	\end{figure}}
\textcolor{Reviewer4}{The homogenized material tensor for the depicted unit cell is determined as follows:
	\begin{center}
	$C^*_{n=12}=\begin{bmatrix}
	22665&4967&14366&-328&-66&328\\
	4967&13396&5968&-287&-65&-50\\
	14366&5968&23035&-7&84&300\\
	-328&-287&-7&5351&-80&-9\\
	-66&-65&84&-80&6280&5\\
	328&-50&300&-9&5&13120\\
	\end{bmatrix}
	$
	\end{center}}
	\textcolor{Reviewer4}{The depicted tensor suggests that the octet-truss unit cell possesses the orthotropic material symmetry. The off-diagonal entries in lines 4--6 are relatively small and can be treated as zero.}
	\textcolor{Reviewer4}{When all of the unit cells are considered, w}e can observe that the mean value of the homogenized numerical solution deviates by 5.3\% from the one determined by using DNS tensile test on the CT scan, but - due to the large geometrical variations in the as-manufactured specimen - the octet-truss unit cells provide a considerable standard deviation from the mean value as shown in~\cref{fig:allTestStressStrain} \textcolor{Reviewer4}{and in~\cref{tab:ComputationalResourcesHomogenization}. The resulting interval of confidence includes both experimental and CT-based DNS simulation results.  The homogenization with FCM is a less memory expensive operation than homogenization with Voxel-FEM. Furthermore, it does not require high-performance computing resources in contrast to the DNS simulation (see~\cref{tab:ComputationalResourcesHomogenization}).} In light of the above results, we can conclude that also first-order homogenization based on the as-manufactured geometry can be a reliable technique to predict the tensile behavior of AM lattice structures.

		\begin{table}[H]
			\centering
			\begin{tabular}{|c|c|c|c|c|}
				\hline
				\textcolor{Reviewer4}{Homogenization setup} & \textcolor{Reviewer4}{$E^*$,[MPa]} & \textcolor{Reviewer4}{Computational nodes} & \textcolor{Reviewer4}{CPU per node} & \textcolor{Reviewer4}{Wall-clock time} \\
				\hline
				\textcolor{Reviewer4}{ FCM}& \textcolor{Reviewer4}{$13769\pm1942$} & \textcolor{Reviewer4}{1} & \textcolor{Reviewer4}{40} & \textcolor{Reviewer4}{80 min}\\
				\textcolor{Reviewer4}{Voxel-FEM$^\ddagger$}& \textcolor{Reviewer4}{$13804\pm1985$}& \textcolor{Reviewer4}{1} &\textcolor{Reviewer4}{40} & \textcolor{Reviewer4}{82 min}\\
				\hline
			\end{tabular}
			\caption{\textcolor{Reviewer4}{Comparison of the results and the computational resources for numerical homogenization ($^\ddagger$ - memory extensive computation).}}
			\label{tab:ComputationalResourcesHomogenization}
		\end{table}

\paragraph{{CAD-based direct numerical simulation of a tensile test \newline}}
	As a next step, the same numerical studies are performed on the original CAD model used for specimen printing. For the direct numerical simulation, \textcolor{Reviewer4}{quadratic solid tetrahedral} elements are chosen to perform analysis in Ansys\textsuperscript{\textregistered}. The converged solution shown in~\cref{tab::YoungsModulusNumerics} is achieved with $10\,954\,356$ DOFs. The total number of DOFs, in this case, is considerably lower than the one used for the CT-based computations. This arises due to the geometrical complexity of the considered CT scan. In the as-manufactured geometry, many more small features are present, such as, e.g., overhangs as indicated in~\cref{fig:PrintedSpecimensComparison2}. Thus, a higher resolution is required to capture the mechanical contribution of such geometrical details to an overall part behavior.

%	\begin{figure}[H]
%	\captionsetup[subfigure]{labelformat=empty}
%	\centering
%	\subfloat[(a) Displacement field ]{\includegraphics[scale=0.1,trim={0cm 0cm 41cm 0cm},clip]{\graphDir/Ux_008.png}}
%	\\
%	\subfloat[(b) Von Mises stress distribution]{\includegraphics[scale=0.1,trim={0cm 0cm 41cm 0cm},clip]{\graphDir/Mises_008.png}}
%	\caption{CAD-based numerical analysis: Representative displacement and stress fields in the middle part of the tensile specimen.}
%	\label{fig::IdealDisplacementStress}
%	\end{figure}
	
	As expected from the higher porosity value measured in~\cref{subsec:geometricalComparison}, the ideal CAD model delivers a much smaller Young's modulus than the experimental one. This relates to the geometrical differences of the as-manufactured and as-designed specimens presented in~\cref{subsec:geometricalComparison}.
	This already demonstrates that the high discrepancy between as-manufactured and as-designed porosity values lead to an inaccurate prediction of Young's modulus based on CAD geometries (see \cref{tab::YoungsModulusNumerics}). The values of Young's moduli are even further apart than the difference between as-manufactured and as-designed porosity. 
	
\paragraph{{CAD-based homogenization with an equivalent porosity \newline}}
	To investigate in detail the observed discrepancy among CAD-based results and experimental measurements, a further study on the CAD geometry is undertaken. The geometrical features of the unit cell depicted in~\cref{fig:unitCellOctet} are varied linearly to achieve different porosity states $\phi$. Such a study can be performed using different approaches. The main geometrical feature of the considered unit cell is the strut diameter. In~\cref{fig:unitCellOctet} all struts horizontal to a printing direction are $0.8$ mm, while the struts inclined to the printing direction are $0.4$ mm. Thus, the dimensions of the struts could be varied freely to achieve different porosity values. However, to reduce the dimension of the undertaken study, both strut diameters are increased by the same increment, e.g., $0.2$ mm. In particular, the diameter of the smallest struts is increased from $0.4$ to $1.0$ while the largest diameter - from $0.8$ to $1.4$. 
	
\begin{figure}[H]
		\captionsetup[subfigure]{labelformat=empty}
		\centering
		\includegraphics[height=0.4\textheight]{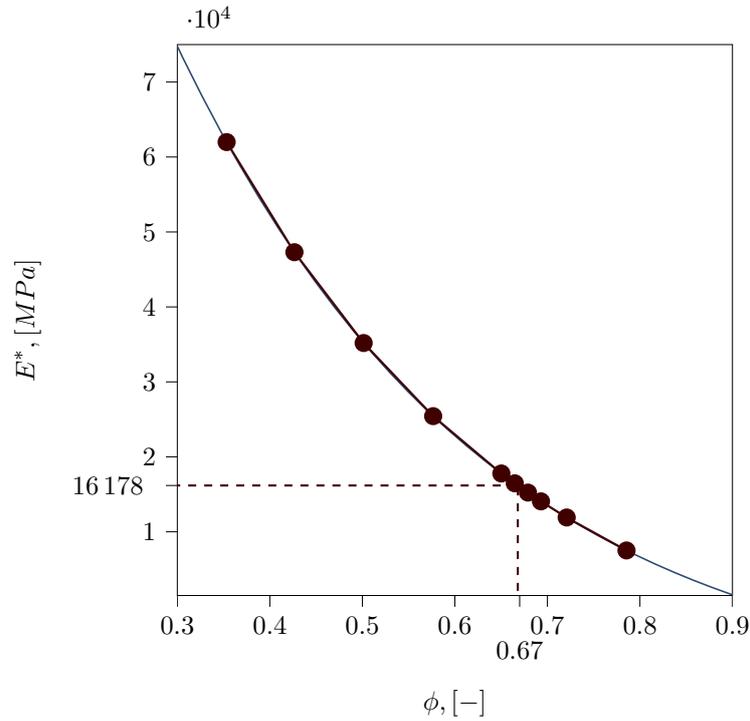}
		\caption{ Porosity study of the ideal CAD-based unit cell using the Finite Cell Method.}
		\label{fig:unitCellStudy}
	\end{figure}

	Then, the first-order homogenization technique described in \cref{sec:homogenization} in combination with the Finite Cell Method is applied to evaluate the effective Young's modulus based on the modified CAD models. The dependency of the homogenized Young's modulus $E^*$ on the porosity is depicted in~\cref{fig:unitCellStudy}. As expected from the literature, a change in the porosity does not cause the same change in the Youngs modulus. Instead, the obtained relation $E^*(\phi)$ seems to be indirectly proportional, but its exact determination is too elaborate to be predicted analytically.

	Under the assumption that the overall porosity is the only determining factor for Young's modulus, an experimentally determined porosity of $0.668$ delivers a value of $16\,178$ MPa. This is much higher than the results of CT-based numerical analysis ($13\,081$ MPa) as well as the experimentally measured value of $12\,533$ MPa. Therefore, the porosity can not be the only determining factor.

\begin{figure}[H]
		\centering
		\includegraphics[scale=0.25,trim={0cm 0cm 0cm 0cm},clip]{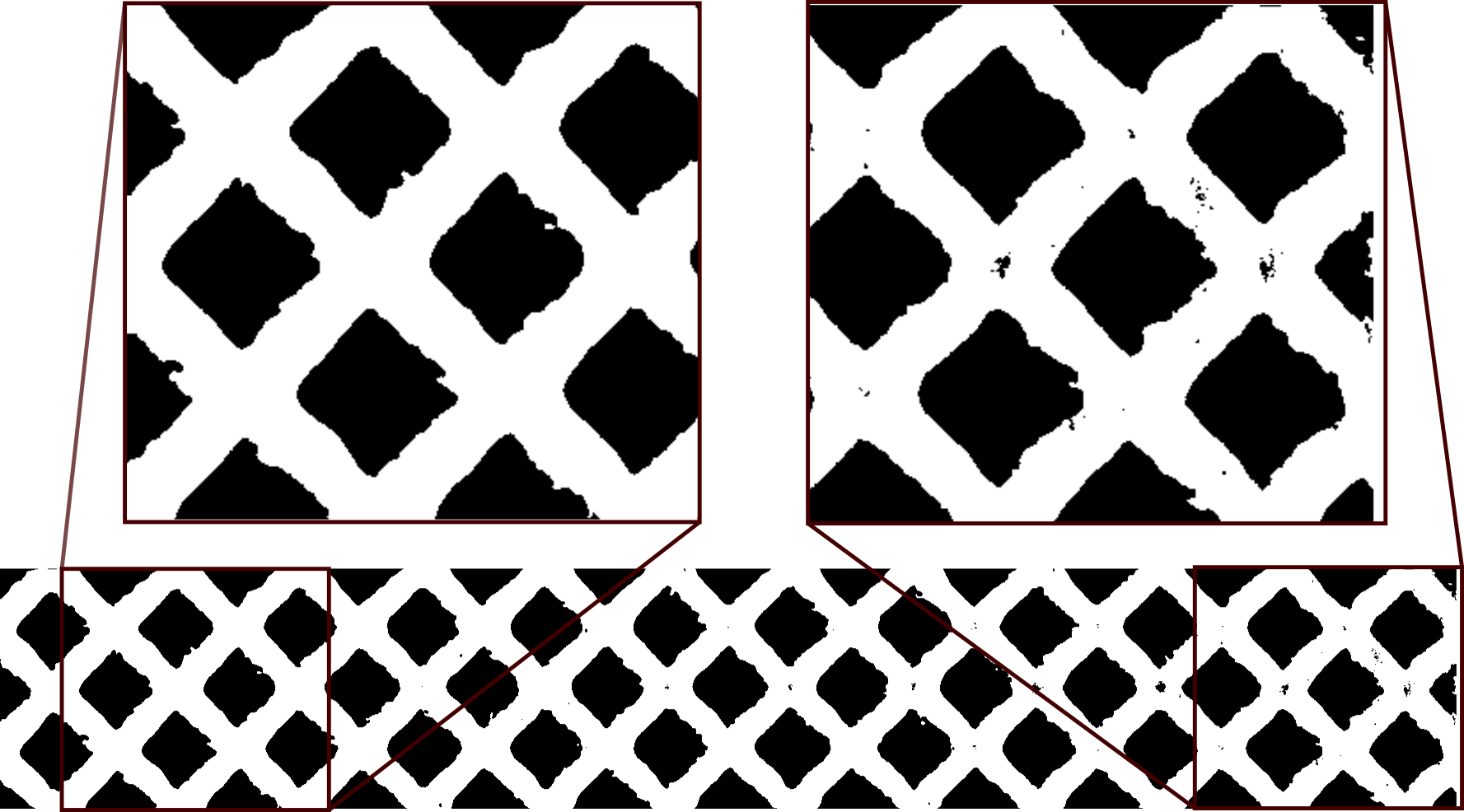}
		\caption{ Representative slice of the CT scan with a zoom on arising inhomogeneities.}
		\label{fig::CTScanZoom}
	\end{figure}

	The strong variability in the printed struts and nodes could explain this difference, raising the importance of the incorporation of the precise as-manufactured geometries into the numerical analysis.	First of all, there is a significant geometrical deviation of the printed struts from as-designed boundaries. Next, the non-homogeneity of material properties of the manufactured component, due, for instance, to voids and inclusions, is present within the printed structure. These inhomogeneities are shown in~\cref{fig::CTScanZoom} on one of the slices of the obtained CT scan.
	
	These topological and geometrical defects introduce an additional level of complexity in the mechanical characterization of AM lattice structures, which cannot be merely modeled with an equivalent CAD-based geometrical model, but it necessarily requires a numerical approach (DNS or homogenization) able to take into account the actual geometry of the manufactured part. The presented numerical framework naturally incorporates the as-manufactured components, thus considering the arising defects.

	\begin{table}[H]
		\centering
		\caption{Comparison of experimentally and numerically determined Young's modulus of the octet-truss specimen.}
		\label{tab::YoungsModulusNumerics}
		\begin{tabular}{|c|c|c|c|}
			\hline
			Specimen &  CAD-based $E$, [MPa]& Experimental $E$, [MPa]&CT-based $E$, [MPa]\\\hline
			Tensile octet-truss & 7 356  & 12 533 $\pm$ 751  & 13 081  \\\hline
		\end{tabular}
	\end{table}

	\section{Conclusions} \label{sec:conclusion}
{
	The incorporation of the process-induced defects of additively manufactured lattice structures in the numerical analysis can be very challenging, time-consuming, and computationally costly. In this contribution, we have shown a natural and efficient method to incorporate as-manufactured geometries in an image-to-mechanical-characterization workflow (indicated in green in~\cref{fig::workflow}). The numerical results obtained via the CT-based numerical analysis utilizing FCM have shown an excellent agreement with the experimental values for the tensile behavior. 
	
	Furthermore, we have undertaken a comprehensive study on the geometrical and mechanical differences between as-designed and as-manufactured specimens. The 3D printed octet-truss lattice structures always possess a larger macroscopic porosity in their designed as compared to their printed version. This deviation correlates to a substantial increase of Young's modulus in the investigated parts. 
	
	From the presented results, we can conclude that the FCM direct numerical simulations based on as-manufactured geometry present a superior capability to predict experimental evidence compared to mesh-conforming DNS performed on the as-designed geometry. Moreover, we have shown that first-order homogenization using FCM directly on CT scan data images is an accurate model of the tensile behavior of octet-truss lattice structures, whereas the same kind of analysis carried out using classical FEA on the CAD model fails to recover the experimental values even when a 3D virtual model with equivalent porosity is used.   
	
	\textcolor{Reviewer1}{ In this paper, we have demonstrated the numerical workflow on an example of specific geometry. However, the proposed methodology is flexible with respect to the characteristic size of the considered parts. First, when a structure with a smaller microstructure is considered, the determining factor for the quality of the achieved numerical results is the resolution of the acquired CT images. The smaller the manufactured components' geometrical features, the finer should be the recorded computed tomography. It is important to note that for microscopic scales, i.e., for features in the range of microns, several problems might occur. Besides the fact that such structures are difficult to manufacture and likewise difficult to test mechanically, imagining problems arise as well. For example, a wide range of metal artifacts can appear in the CT images. These make difficult a reliable distinction between partially melted but not load-bearing powder and the solidified load-bearing material. As a remedy to this problem, a combination of the proposed workflow with deep learning segmentation algorithms can be used. Such an approach was proposed by the authors in~\cite{Korshunova2020}. Second, when much larger parts with complex microstructure are considered, this workflow's limiting factor is the available computational resources. Undoubtedly, the higher the scale separation between the component and its microstructure, the more computationally demanding the numerical analysis becomes. Indeed, the proposed hybrid parallelization technique allows for the computation of industrial-scale examples. Nevertheless, when the scale separation becomes too high, the proposed numerical homogenization technique can be used on the smaller scale and mapped back to the component size which would further push the computational boundaries.}
	
	To conclude, the proposed workflow is able to reduce the gap between design and analysis in the field of AM product simulation by a combination of several efficient methodologies. Future work will address the generation of statistically similar computed tomography images and their incorporation in the AM product simulation workflow to characterize the possible effects of the geometrical defects on the mechanical behavior of the final parts (shown in dashed green in~\cref{fig::workflow}). \textcolor{Reviewer1}{To this end, CT images of as-manufactured parts can be analyzed without a pre-defined set of the occurring topological and geometrical deviations. Instead, the process-induced defects, such as, e.g., internal porosity, strut oversizing, or strut connectivity, can be considered as a whole. Such a holistic approach allows for an estimation of the variability of the mechanical properties to the full extent. Furthermore, the generation of stochastic geometries could help investigate the influence of the process parameters on the manufactured parts. In particular, combining multiple CT images in such a workflow could provide an insight into which impact the process parameters, e.g., the printing direction or the laser power, have on the manufactured lattices' microstructure.} This step complements the proposed method to enhance the possibility of the natural incorporation of the printed AM geometries into the numerical analysis.
}

%% Acknowledgements -------------------------------  
\section*{Acknowledgements} 
We gratefully acknowledge the support of Deutsche Forschungsgemeinschaft (DFG) through the project 414265976 – TRR 277 C-01 and TUM International Graduate School of Science and Engineering (IGSSE), GSC 81. The authors also gratefully acknowledge the Gauss Centre for Supercomputing e.V. (www.gauss-centre.eu) for funding this project by providing computing time on the Linux Cluster CoolMUC-2 and on the GCS Supercomputer SuperMUC-NG at Leibniz Supercomputing Centre (www.lrz.de). This work was partially supported by the Italian Minister of University and Research through the MIUR-PRIN projects "A BRIDGE TO THE FUTURE: Computational methods, innovative applications, experimental validations of new materials and technologies” (No. 2017L7X3CS) and "XFAST-SIMS" (no. 20173C478N). The authors would like to acknowledge the project "MADE4LO - Metal ADditivE for LOmbardy" (No. 240963) within the POR FESR 2014-2020 program. We also kindly acknowledge Eng. Alberto Cattenone and Prof. Stefania Marconi of the 3DMetal laboratory of the Department of Civil Engineering and Architecture of the University of Pavia for providing facilities for additive manufacturing and experimental testing (http://www-4.unipv.it/3d/laboratories/3dmetalunipv/). The authors gratefully acknowledge Giorgio Vattasso from LaborMet Due (http://www.labormetdue.it/) for his technical support in obtaining CT scan images. Finally, we acknowledge Academy of Finland through the project Adaptive isogeometric methods for thin-walled structures (decision number 304122) as well as the August-Wilhelm Scheer Visiting Professors Program established by TUM International Center and funded by the German Excellence Initiative.  

\newpage
%% References -------------------------------------
\bibliographystyle{apalike}
 \bibliography{library}

\begin{thebibliography}{}

\bibitem[Al-Saedi et~al., 2018]{Saedi2018}
Al-Saedi, D., Masood, S., Faizan-Ur-Rab, M., Alomarah, A., and Ponnusamy, P.
  (2018).
\newblock Mechanical properties and energy absorption capability of
  functionally graded {F2BCC} lattice fabricated by {SLM}.
\newblock {\em Materials \& Design}, 144:32--44.

\bibitem[{ASTM International}, 2017]{ASTME111}
{ASTM International} (2017).
\newblock {\em {ISO / ASTM52900-15}: Standard Terminology for Additive
  Manufacturing – General Principles – Terminology,}.
\newblock West Conshohocken, PA: American Society for Testing and Materials.

\bibitem[Babuska, 1973]{Babuska1973}
Babuska, I. (1973).
\newblock The {{Finite Element Method}} with {{Penalty}}.
\newblock {\em Mathematics of Computation}, 27(122):221--228.

\bibitem[Bagheri et~al., 2017]{Bagheri2017}
Bagheri, Z.~S., Melancon, D., Liu, L., Johnston, R.~B., and Pasini, D. (2017).
\newblock Compensation strategy to reduce geometry and mechanics mismatches in
  porous biomaterials built with {{Selective Laser Melting}}.
\newblock {\em Journal of the Mechanical Behavior of Biomedical Materials},
  70:17--27.

\bibitem[Buchanan and Gardner, 2019]{Buchanan2019}
Buchanan, C. and Gardner, L. (2019).
\newblock Metal {3D} printing in construction: A review of methods, research,
  applications, opportunities and challenges.
\newblock {\em Engineering Structures}, 180:332--348.

\bibitem[Campoli et~al., 2013]{Campoli2013}
Campoli, G., Borleffs, M.~S., Amin~Yavari, S., Wauthle, R., Weinans, H., and
  Zadpoor, A.~A. (2013).
\newblock Mechanical properties of open-cell metallic biomaterials manufactured
  using additive manufacturing.
\newblock {\em Materials \& Design}, 49:957--965.

\bibitem[Dallago et~al., 2018]{Dallago2018a}
Dallago, M., Fontanari, V., Torresani, E., Leoni, M., Pederzolli, C., Potrich,
  C., and Benedetti, M. (2018).
\newblock Fatigue and biological properties of ti-6al-4v eli cellular
  structures with variously arranged cubic cells made by selective laser
  melting.
\newblock {\em Journal of the Mechanical Behavior of Biomedical Materials},
  78:381--394.

\bibitem[Dallago et~al., 2017]{Dallago2017}
Dallago, M., Fontanari, V., Winiarski, B., Zanini, F., Carmignato, S., and
  Benedetti, M. (2017).
\newblock Fatigue properties of ti6al4v cellular specimens fabricated via slm:
  Cad vs real geometry.
\newblock {\em Procedia Structural Integrity}, 7:116--123.
\newblock 3rd International Symposium on Fatigue Design and Material Defects,
  FDMD 2017.

\bibitem[Dallago et~al., 2021]{Dallago2021}
Dallago, M., Raghavendra, S., Luchin, V., Zappini, G., Pasini, D., and
  Benedetti, M. (2021).
\newblock The role of node fillet, unit-cell size and strut orientation on the
  fatigue strength of {Ti-6Al-4V} lattice materials additively manufactured via
  laser powder bed fusion.
\newblock {\em International Journal of Fatigue}, 142:105946.

\bibitem[Dallago et~al., 2019]{Dallago2019a}
Dallago, M., Winiarski, B., Zanini, F., Carmignato, S., and Benedetti, M.
  (2019).
\newblock On the effect of geometrical imperfections and defects on the fatigue
  strength of cellular lattice structures additively manufactured via selective
  laser melting.
\newblock {\em International Journal of Fatigue}, 124:348 -- 360.

\bibitem[Dong et~al., 2017]{Dong2017}
Dong, G., Tang, Y., and Zhao, Y.~F. (2017).
\newblock A survey of modeling of lattice structures fabricated by additive
  manufacturing.
\newblock {\em Journal of Mechanical Design}, 139:32--38.

\bibitem[Dong et~al., 2019]{Dong2019}
Dong, Z., Liu, Y., Li, W., and Liang, J. (2019).
\newblock Orientation dependency for microstructure, geometric accuracy and
  mechanical properties of selective laser melting {{AlSi10Mg}} lattices.
\newblock {\em Journal of Alloys and Compounds}, 791:490--500.

\bibitem[{du Plessis} et~al., 2018]{Duplesis2018}
{du Plessis}, A., Sperling, P., Beerlink, A., Tshabalala, L., Hoosain, S.,
  Mathe, N., and {le Roux}, S.~G. (2018).
\newblock Standard method for microct-based additive manufacturing quality
  control 2: Density measurement.
\newblock {\em MethodsX}, 5:1117--1123.

\bibitem[{du Plessis} et~al., 2020]{Duplesis2020}
{du Plessis}, A., Yadroitsava, I., and Yadroitsev, I. (2020).
\newblock Effects of defects on mechanical properties in metal additive
  manufacturing: A review focusing on {X-ray} tomography insights.
\newblock {\em Materials \& Design}, 187:108385.

\bibitem[D{\"u}ster et~al., 2017]{Duster2017}
D{\"u}ster, A., Rank, E., and Szab{\'o}, B.~A. (2017).
\newblock The p-version of the finite element method and finite cell methods.
\newblock In Stein, E., Borst, R., and Hughes, T. J.~R., editors, {\em
  Encyclopedia of {{Computational}} Mechanics}, volume~2, pages 1--35. {John
  Wiley \& Sons}, {Chichester, West Sussex}.

\bibitem[Elhaddad et~al., 2018]{Elhaddad2018}
Elhaddad, M., Zander, N., Bog, T., Kudela, L., Kollmannsberger, S., Kirschke,
  J.~S., Baum, T., Ruess, M., and Rank, E. (2018).
\newblock Multi-level hp-finite cell method for embedded interface problems
  with application in biomechanics.
\newblock {\em International Journal for Numerical Methods in Biomedical
  Engineering}, 34(4):e2951.

\bibitem[Elhaddad et~al., 2015]{Elhaddad2015}
Elhaddad, M., Zander, N., Kollmannsberger, S., Shadavakhsh, A., N{\"u}bel, V.,
  and Rank, E. (2015).
\newblock Finite {{Cell Method}}: {{High}}-{{Order Structural Dynamics}} for
  {{Complex Geometries}}.
\newblock {\em International Journal of Structural Stability and Dynamics},
  15(7):1540018.

\bibitem[Geng et~al., 2019]{Geng2019}
Geng, L., Wu, W., Sun, L., and Fang, D. (2019).
\newblock Damage characterizations and simulation of selective laser melting
  fabricated {3D} re-entrant lattices based on in-situ {CT} testing and
  geometric reconstruction.
\newblock {\em International Journal of Mechanical Sciences}, 157-158:231 --
  242.

\bibitem[Gross and Seelig, 2017]{Gross2017}
Gross, D. and Seelig, T. (2017).
\newblock {\em Fracture {{Mechanics}}: {{With}} an {{Introduction}} to
  {{Micromechanics}}}.
\newblock Mechanical {{Engineering Series}}. {Springer International
  Publishing}.

\bibitem[Hazanov and Huet, 1994]{Hazanov1994}
Hazanov, S. and Huet, C. (1994).
\newblock Order relationships for boundary conditions effect in heterogeneous
  bodies smaller than the representative volume.
\newblock {\em Journal of the Mechanics and Physics of Solids},
  42(12):1995--2011.

\bibitem[Heinze et~al., 2017]{Heinze2017}
Heinze, S., Bleistein, T., Düster, A., Diebels, S., and Jung, A. (2017).
\newblock Experimental and numerical investigation of single pores for
  identification of effective metal foams properties.
\newblock {\em Journal of Applied Mathematics and Mechanics}, 98:682--695.

\bibitem[{ISO/ASTM52900-15}, 2015]{ASTM52900}
{ISO/ASTM52900-15} (2015).
\newblock {\em {ASTM E 111-17}: Standard Test Method for Young’s Modulus,
  Tangent Modulus, and Chord Modulus}.
\newblock West Conshohocken, PA: American Society for Testing and Materials.

\bibitem[Jomo et~al., 2017]{Jomo2016}
Jomo, J., Zander, N., Elhaddad, M., {\"O}zcan, A.~I., Kollmannsberger, S.,
  Mundani, R.-P., and Rank, E. (2017).
\newblock Parallelization of the multi-level hp-adaptive finite cell method.
\newblock {\em Computers and Mathematics with Applications}, 74(1):126--142.

\bibitem[Jomo et~al., 2019]{Jomo2019}
Jomo, J.~N., {de Prenter}, F., Elhaddad, M., D'Angella, D., Verhoosel, C.~V.,
  Kollmannsberger, S., Kirschke, J.~S., N{\"u}bel, V., {van Brummelen}, E.~H.,
  and Rank, E. (2019).
\newblock Robust and parallel scalable iterative solutions for large-scale
  finite cell analyses.
\newblock {\em Finite Elements in Analysis and Design}, 163:14--30.

\bibitem[Korshunova et~al., 2020]{Korshunova2020}
Korshunova, N., Jomo, J., L{\'e}k{\'o}, G., Reznik, D., Bal{\'a}zs, P., and
  Kollmannsberger, S. (2020).
\newblock Image-based material characterization of complex microarchitectured
  additively manufactured structures.
\newblock {\em Computers \& Mathematics with Applications}, 80(11):2462 --
  2480.

\bibitem[Lei et~al., 2019]{Lei2019}
Lei, H., Li, C., Meng, J., Zhou, H., Liu, Y., Zhang, X., Wang, P., and Fang, D.
  (2019).
\newblock Evaluation of compressive properties of {SLM}-fabricated multi-layer
  lattice structures by experimental test and $\mu$-{CT}-based finite element
  analysis.
\newblock {\em Materials \& Design}, 169:107685.

\bibitem[Lietaert et~al., 2018]{Lietaert2018}
Lietaert, K., Cutolo, A., Boey, D., and Van~Hooreweder, B. (2018).
\newblock Fatigue life of additively manufactured ti6al4v scaffolds under
  tension-tension, tension-compression and compression-compression fatigue
  load.
\newblock {\em Scientific Reports}, 8:4957.

\bibitem[Liu et~al., 2017]{Liu2017}
Liu, L., Kamm, P., {Garc{\'i}a-Moreno}, F., Banhart, J., and Pasini, D. (2017).
\newblock Elastic and failure response of imperfect three-dimensional metallic
  lattices: The role of geometric defects induced by {{Selective Laser
  Melting}}.
\newblock {\em Journal of the Mechanics and Physics of Solids}, 107:160--184.

\bibitem[Lozanovski et~al., 2019]{Lozanovski2019}
Lozanovski, B., Leary, M., Tran, P., Shidid, D., Qian, M., Choong, P., and
  Brandt, M. (2019).
\newblock Computational modelling of strut defects in {SLM} manufactured
  lattice structures.
\newblock {\em Materials \& Design}, 171:107671.

\bibitem[Maconachie et~al., 2019]{Maconachie2019}
Maconachie, T., Leary, M., Lozanovski, B., Zhang, X., Qian, M., Faruque, O.,
  and Brandt, M. (2019).
\newblock {{SLM}} lattice structures: {{Properties}}, performance, applications
  and challenges.
\newblock {\em Materials \& Design}, 183:108137.

\bibitem[Mazur et~al., 2017]{Mazur2017}
Mazur, M., Leary, M., McMillan, M., Sun, S., Shidid, D., and Brandt, M. (2017).
\newblock 5 - mechanical properties of ti6al4v and alsi12mg lattice structures
  manufactured by selective laser melting (slm).
\newblock In Brandt, M., editor, {\em Laser Additive Manufacturing}, Woodhead
  Publishing Series in Electronic and Optical Materials, pages 119--161.
  Woodhead Publishing.

\bibitem[Melancon et~al., 2017]{Melancon2017}
Melancon, D., Bagheri, Z., Johnston, R., Liu, L., Tanzer, M., and Pasini, D.
  (2017).
\newblock Mechanical characterization of structurally porous biomaterials built
  via additive manufacturing: experiments, predictive models, and design maps
  for load-bearing bone replacement implants.
\newblock {\em Acta Biomaterialia}, 63:350--368.

\bibitem[Nemat-Nasser et~al., 2013]{Nemat2013}
Nemat-Nasser, S., Hori, M., and Achenbach, J. (2013).
\newblock {\em Micromechanics: Overall Properties of Heterogeneous Materials}.
\newblock ISSN. Elsevier Science.

\bibitem[Ngo et~al., 2018]{Ngo2018}
Ngo, T.~D., Kashani, A., Imbalzano, G., Nguyen, K. T.~Q., and Hui, D. (2018).
\newblock Additive manufacturing ({{3D}} printing): {{A}} review of materials,
  methods, applications and challenges.
\newblock {\em Composites Part B: Engineering}, 143:172--196.

\bibitem[Nguyen et~al., 2012]{Nguyen2012a}
Nguyen, V.-D., B{\'e}chet, E., Geuzaine, C., and Noels, L. (2012).
\newblock Imposing periodic boundary condition on arbitrary meshes by
  polynomial interpolation.
\newblock {\em Computational Materials Science}, 55:390--406.

\bibitem[Pahr and Zysset, 2008]{Pahr2008}
Pahr, D.~H. and Zysset, P.~K. (2008).
\newblock Influence of boundary conditions on computed apparent elastic
  properties of cancellous bone.
\newblock {\em Biomechanics and Modeling in Mechanobiology}, 7(6):463--476.

\bibitem[Parvizian et~al., 2007]{Parvizian2007}
Parvizian, J., D{\"u}ster, A., and Rank, E. (2007).
\newblock Finite cell method.
\newblock {\em Computational Mechanics}, 41(1):121--133.

\bibitem[Pasini and Guest, 2019]{Pasini2019}
Pasini, D. and Guest, J.~K. (2019).
\newblock Imperfect architected materials: {{Mechanics}} and topology
  optimization.
\newblock {\em MRS Bulletin}, 44(10):766--772.

\bibitem[Portela et~al., 2018]{Portela2018}
Portela, C.~M., Greer, J.~R., and Kochmann, D.~M. (2018).
\newblock Impact of node geometry on the effective stiffness of non-slender
  three-dimensional truss lattice architectures.
\newblock {\em Extreme Mechanics Letters}, 22:138 -- 148.

\bibitem[Płatek et~al., 2020]{Platek2020}
Płatek, P., Sienkiewicz, J., Janiszewski, J., and Jiang, F. (2020).
\newblock {}investigations on mechanical properties of lattice structures with
  different values of relative density made from {316L} by {Selective} {Laser}
  {Melting} ({SLM}).
\newblock page 2204.

\bibitem[Rashed et~al., 2016]{Rashed2016}
Rashed, M., Ashraf, M., Mines, R., and Hazell, P.~J. (2016).
\newblock Metallic microlattice materials: A current state of the art on
  manufacturing, mechanical properties and applications.
\newblock {\em Materials \& Design}, 95:518 -- 533.

\bibitem[Refai et~al., 2020a]{Refai2020b}
Refai, K., Brugger, C., Montemurro, M., and Saintier, N. (2020a).
\newblock An experimental and numerical study of the high cycle multiaxial
  fatigue strength of titanium lattice structures produced by selective laser
  melting (slm).
\newblock {\em International Journal of Fatigue}, 138:105623.

\bibitem[Refai et~al., 2020b]{Refai2020a}
Refai, K., Montemurro, M., Brugger, C., and Saintier, N. (2020b).
\newblock Determination of the effective elastic properties of titanium lattice
  structures.
\newblock {\em Mechanics of Advanced Materials and Structures},
  27(23):1966--1982.

\bibitem[Renishaw-PLC, 2020]{ManualReinshaw}
Renishaw-PLC (2020).
\newblock {\em Data sheet: SS 316L-0407 powder for additive manufacturing}.
\newblock Renishaw-PLC.

\bibitem[Sanaei et~al., 2019]{Sanae2019}
Sanaei, N., Fatemi, A., and Phan, N. (2019).
\newblock Defect characteristics and analysis of their variability in metal
  {L-PBF} additive manufacturing.
\newblock {\em Materials \& Design}, 182:108091.

\bibitem[Shidid et~al., 2016]{Shidid2016}
Shidid, D., Leary, M., Choong, P., and Brandt, M. (2016).
\newblock Just-in-time design and additive manufacture of patient-specific
  medical implants.
\newblock {\em Physics Procedia}, 83:4 -- 14.

\bibitem[Sing et~al., 2018]{Sing2018}
Sing, S.~L., Wiria, F.~E., and Yeong, W.~Y. (2018).
\newblock Selective laser melting of lattice structures: A statistical approach
  to manufacturability and mechanical behavior.
\newblock {\em Robotics and Computer-Integrated Manufacturing}, 49:170 -- 180.

\bibitem[Suquet, 1985]{Suquet1985}
Suquet, P. (1985).
\newblock {\em Elements of Homogenization for Inelastic Solid Mechanics}.

\bibitem[Tancogne-Dejean and Mohr, 2018]{Tancogne2018}
Tancogne-Dejean, T. and Mohr, D. (2018).
\newblock Elastically-isotropic truss lattice materials of reduced plastic
  anisotropy.
\newblock {\em International Journal of Solids and Structures}, 138:24 -- 39.

\bibitem[Tian et~al., 2019]{Tian2019}
Tian, W., Qi, L., Chao, X., Liang, J., and Fu, M. (2019).
\newblock Periodic boundary condition and its numerical implementation
  algorithm for the evaluation of effective mechanical properties of the
  composites with complicated micro-structures.
\newblock {\em Composites Part B: Engineering}, 162:1 -- 10.

\bibitem[Vigliotti et~al., 2014]{Vigliotti2014}
Vigliotti, A., Deshpande, V.~S., and Pasini, D. (2014).
\newblock Non linear constitutive models for lattice materials.
\newblock {\em Journal of the Mechanics and Physics of Solids}, 64:44 -- 60.

\bibitem[Wang et~al., 2019]{Wang2019}
Wang, P., Lei, H., Zhu, X., Chen, H., and Fang, D. (2019).
\newblock Influence of manufacturing geometric defects on the mechanical
  properties of {AlSi10Mg} alloy fabricated by selective laser melting.
\newblock {\em Journal of Alloys and Compounds}, 789:852 -- 859.

\bibitem[Yan et~al., 2012]{Yan2012}
Yan, C., Hao, L., Hussein, A., and Raymont, D. (2012).
\newblock Evaluations of cellular lattice structures manufactured using
  selective laser melting.
\newblock {\em International Journal of Machine Tools and Manufacture},
  62:32--38.

\bibitem[Yang et~al., 2017]{YangBook2017}
Yang, L., Hsu, K., Baughman, B., Godfrey, D., Medina, F., Menon, M., and
  Wiener, S. (2017).
\newblock {\em Additive Manufacturing of Metals: The Technology, Materials,
  Design and Production}.
\newblock Springer International Publishing.

\bibitem[Yang et~al., 2012]{Yang2012a}
Yang, Z., Ruess, M., Kollmannsberger, S., D{\"u}ster, A., and Rank, E. (2012).
\newblock An efficient integration technique for the voxel-based finite cell
  method.
\newblock {\em International Journal for Numerical Methods in Engineering},
  91(5):457--471.

\bibitem[Zhou et~al., 2019]{Zhou2019}
Zhou, H., Zhang, X., Zeng, H., Yang, H., Lei, H., Li, X., and Wang, Y. (2019).
\newblock Lightweight structure of a phase-change thermal controller based on
  lattice cells manufactured by slm.
\newblock {\em Chinese Journal of Aeronautics}, 32(7):1727 -- 1732.

\bibitem[Zohdi and Wriggers, 2004]{Zohdi2004}
Zohdi, T. and Wriggers, P. (2004).
\newblock {\em An Introduction to Computational Micromechanics}.
\newblock Lecture Notes in Applied and Computational Mechanics. Springer Berlin
  Heidelberg.

\end{thebibliography}

\end{document}